\newcommand{\diracslash}[1]{#1\llap{/\kern2pt}}

\newcommand{\be}{\begin{equation}}
\newcommand{\ee}{\end{equation}}
\newcommand{\bea}{\begin{eqnarray}}
\newcommand{\eea}{\end{eqnarray}}
\newcommand{\ba}[1]{\begin{array}{#1}}
\newcommand{\ea}{\end{array}}

\newcommand{\bt}{\begin{tabular}}
\newcommand{\et}{\end{tabular}}

\newcommand{\beas}{\begin{eqnarray*}}
\newcommand{\eeas}{\end{eqnarray*}}

\documentclass[preprint,prd,aps,floats,nofootinbib,floatfix]{revtex4-1}
\usepackage{graphicx}
\begin{document}

\title {Decay widths of bottomonium states
in matter -- a field theoretic model for composite hadrons}
\author{Amruta Mishra}
\email{amruta@physics.iitd.ac.in}
\affiliation{Department of Physics, Indian Institute of Technology, Delhi,
Hauz Khas, New Delhi -- 110 016, India}

\author{S.P. Misra}
\email{misrasibaprasad@gmail.com}
\affiliation{Institute of Physics, Bhubaneswar -- 751005, India} 

\begin{abstract}
We compute the in-medium partial decay widths of the bottomonium 
states to open bottom mesons ($B\bar B$) using a field 
theoretical model for composite hadrons with quark constituents. 
These decay widths are calculated by using the explicit constructions 
for the bottomonium states and the open bottom mesons ($B$ and $\bar B$),
and, the quark antiquark pair creation term of the free Dirac Hamiltonian 
written in terms of the constituent quark field operators.
These decay widths in the hadronic medium are calculated 
as arising from the mass 
modifications of the bottomonium states and the $B$ and $\bar B$ 
mesons, obtained in a chiral effective model. The decay amplitude 
in the present model is multiplied with a strength parameter for 
the light quark pair creation, which is fitted from the observed 
vacuum partial decay width of the bottomonium state, $\Upsilon (4S)$ 
to $B\bar B$. The effects of the isospin asymmetry, 
the strangeness fraction of the hadronic matter 
on the decay widths, arising due to the mass modifications 
due to these effects, have also been studied. There is observed
to be appreciable effects from density, and 
the effects from isospin asymmetry on the parital decay widths 
of $\Upsilon \rightarrow B\bar B$ are observed to be quite 
pronounced at high densities. These effects should show up in  
the asymmetric heavy ion collisions in Compressed baryonic matter
(CBM) experiments planned at the future facility at FAIR.
The study of the $\Upsilon$ states will, however, require access
to  energies higher than the energy regime planned at the CBM experiment.
The density effects on the decay widths of the bottomonium states
should show up in the production of these states, as well as,
in dilepton spectra at the Super Proton Synchrotron (SPS) energies.

\end{abstract}

\maketitle

\def\bfm#1{\mbox{\boldmath $#1$}}
\def\bfs#1{\mbox{\bf #1}}

\section{Introduction}
The study of medium modifications of properties of hadrons is a topic
of research which has attracted a lot of attention in recent years
in strong interaction physics, in particular because of its relevance 
to the heavy ion collision experiments. Matter at high temperatures 
and/or densities is produced in ultra-relativistic heavy ion collision 
experiments and the properties of hadrons in such a medium are modified,
consequences of which can show up in the experimental observables  
of these high energy nuclear collisions.

The open charm (bottom) mesons, $D$ ($\bar B$) and $\bar D$ ($B$),
are made up of a heavy charm (bottom) quark (antiquark) 
and a light (u or d) antiquark (quark) and their mass modifications
in the hadronic medium are due to their interaction with
the light quark condensate in the QCD sum rule framework
\cite{haya1,qcdsum08}. 
The in-medium properties of the open charm mesons have been
studied quite extensively by hadronic frameworks, 
e.g., the quark meson coupling (QMC) model 
\cite{qmc0,qmchnm,qmc1,qmc2} as well as 
the coupled channel approach \cite{ltolos,ljhs,mizutani6,mizutani8,HL}. 
Within a hadronic framework using pion exchange \cite{sudoh}, 
a study of the open charm and open bottom mesons
is observed to lead to an attractive interaction of the 
$\bar D$ and $B$ mesons in the nuclear matter,
suggesting that these mesons can form bound states with the atomic nuclei.
The $\bar D$-nucleon interactions have recently been studied 
using a description of the hadrons with quark and antiquark constituents
\cite{dnkrein}, where the field operators of the constituent quarks are 
written in terms of a constituent quark mass, 
which arises from dynamical chiral symmetry breaking 
\cite{dnkrein,hmamspm1,hmamspm2,spmeffpot,spmbothcond,amhm2004}.

In the effective hadronic model, constructed by generalizing
chiral SU(3) model to the charm and bottom sectors, 
the mass modifications of these open charm mesons
\cite{amarindam,amarvind,amcharmdw}
and the open bottom mesons \cite {dpambmeson},
arise due to their interactions with the light hadrons,
namely the baryons (nucleons and hyperons) and the scalar
mesons. On the other hand, the hidden charm and bottom mesons, i.e. 
the charmonium \cite{amcharmdw,leeko,kimlee,charmmass2} and 
bottomonium states \cite{amdpbottomonium}, have the
masses modified in the hadronic medium due to the interactions
with the gluon condensates in the medium.
The gluon condensates of QCD is
mimicked by a scalar dilaton field \cite{charmmass2,amcharmdw},
within the effective hadronic model, 
and the medium modifications of the heavy quarkonium states,
i.e., the charmonium \cite{amcharmdw} and bottomonium states
\cite{amdpbottomonium}, are studied
by medium modification of the dilaton field within the
model. 
Using a field theoretical model
for composite hadrons with quark and antiquark constitutents
 \cite{spm781,spm782,spmdiffscat},
the partial decay widths of the charmonium states to 
$D\bar D$ pair, as well as of the decay $D^*\rightarrow D\pi$,
in matter have been studied 
\cite{chmdwamspmwg}, using the medium modifications of
these hadrons using the effective hadronic model
\cite{amcharmdw}. 
These decay widths were compared with the
partial decay widths using the $^3P_0$ model \cite{friman,amcharmdw}, 
where a light quark aniquark pair is assumed to be created 
in the $^3P_0$ state \cite{yopr1,yopr2,yopr3,barnes}, 
and the light quark (antiquark) combines with the heavy 
charm antiquark (quark) of the decaying charmonium state, 
to produce the $D\bar D$ pair.
In the present work, we study the medium modification 
of the decay widths 
of the bottomonium states to $B\bar B$ pair 
in the strange hadronic medium, due to the mass modifications 
of these hadrons calculated in the effective chiral model
\cite{amdpbottomonium,dpambmeson}. 

The outline of the paper is as follows: In section II, we describe 
briefly the field theoretical model for the 
hadrons with quark constitutents, which is used in the present
work to calculate the partial decay widths of the bottomonium 
states to open bottom mesons ($B\bar B$ pair).
The decay widths are calculated by using explicit constructions 
of the bottomonium states ($\Upsilon (1S)$, $\Upsilon (2S)$,
$\Upsilon (3S)$ and $\Upsilon (4S)$) and the $B$ 
and $\bar B$ mesons in terms of the quark and antiquark constituents.
We then calculate the matrix element of the S-matrix in the lowest order
to compute the decay widths of the bottomonium states to $B\bar B$ 
($B^+ B^-$ or $B^0\bar {B^0}$) pair. 
The matrix element, however, is multiplied 
with a parameter, which is fitted from the observed vacuum decay width
of $\Upsilon (4S) \rightarrow B\bar B$.
In the present work, the partial decay widths for the
decay of the bottomonium states, 
$\Upsilon (NS)$, $N$=1,2,3,4, 
to $B\bar B$,
are calculated using the field theoretic model 
for composite hadrons and their medium modifications
have been studied as arising from the changes in the masses
of these mesons in the hadronic medium.
In section III, we briefly describe the effective hadronic
model, which has been used to investigate the masses of the
open bottom mesons ($B$ and $\bar B$) and of the
$\Upsilon$ states. The in-medium masses of the $B$ and
$\bar B$ mesons in the strange hadronic medium arise
due to their interactions with the baryons and the scalar mesons
\cite{dpambmeson}. On the other hand, the mass modifications
of the bottomonium states 
\cite{amdpbottomonium} airse due to 
the medium modification of the scalar dilaton field,
which is incorporated in the effective hadronic framework,
to simulate the scale symmetry breaking of QCD through
scalar gluon condensate. In section IV, we discuss the results 
obtained in the present investigation. Using the explicit
constructions for the bottomonium states ($\Upsilon (NS)$, N=1,2,3,4)
and using the quark pair creation term of the free Dirac 
Hamiltonian written in terms of the constituent quark field operators,
the decay widths of the bottomonium states to $B\bar B$ pair,
are calculated within the present model.
In section V, we summarize 
the results for the medium modifications of these decay widths,
and discuss possible outlook.

\section{The model for composite hadrons}

The model used in the present work for calculating the partial 
decay widths of the bottomonium states to $B\bar B$, describes the 
hadrons comprising of the quark and antiquark constituents. 
In the present section, we shall very briefly describe the model 
so as to apply the same for investigating these decay widths.

The field operator for a constituent quark for a hadron at rest
at time, t=0, is written as
\begin {eqnarray}
\psi ({\bf x},t=0)
&=&(2\pi)^{-{3}/{2}}{\int \Big [U({\bf k}) q_I ({\bf k})
\exp(i{\bf k} \cdot{\bf x})
+V({\bf k}) \tilde q_I ({\bf k})
\exp(-i{\bf k} \cdot{\bf x})\Big ]} d{\bfs k}\nonumber \\
 & \equiv & Q({\bf x})+\tilde Q({\bf x}).
\label{qx}
\end{eqnarray}
In the above, $q_I ({\bf k})=q_r ({\bf k})u_r$ and 
$ \tilde q_I ({\bf k})=\tilde q_s ({\bf k})v_s$ 
are the two component quark annihilation and antiquark creation 
operators. The operator $q_{r}({\bf k})$ annihilates a quark with spin $r$ 
and momentum ${\bf k}$, whereas, $\tilde q _{s}({\bf k})$
creates an antiquark with spin $s$ and momentum ${\bf k}$,
and these operators satisfy the usual anticommutation relations
\begin{equation}
\{q_{r}({\bf k}),q_{s}({\bf k}')^\dagger\}=
\{\tilde q_{r}({\bf k}),\tilde q_{s}({\bf k}')^\dagger\}=
\delta _{rs} \delta ({\bf k}-{\bf k}').
\end{equation}
In equation (\ref{qx}), $U({\bfs k})$ and $V({\bfs k})$ are given as
\begin{equation}
U({\bfs k})=\left (\begin{array}{c} f(|{{\bf k}}|)\\
{\bfm\sigma}\cdot {\bf k} g(|{{\bf k}}|)\\
\end{array} \right ),\;\;\;\;\;
V({\bfs k})=\left (\begin{array}{c} 
{\bfm\sigma}\cdot {\bf k} g(|{{\bf k}}|)\\
f(|{{\bf k}}|)\\
\end{array} \right ),
\label{ukvk}
\end{equation}
where the functions $f(|{\bf k}|)$ and $g(|{\bf k}|)$ satisfy the constraint
\cite{spm781},
\begin{equation}
f^2+g^2 {\bf k}^2=1,
\label{fgk}
\end{equation}
as obtained from the equal time anticommutation relation 
for the four-component Dirac field operators. 
These functions, for the case of free Dirac field
of mass $M$, are given as,
\begin{equation}
f(|{\bf k}|)=\left ( \frac{k_0 +M}{2 k_0}\right )^{1/2},\;\;\;\; 
g(|{\bf k}|)=\left ( \frac{1}{2 k_0 (k_0+M)}\right )^{1/2},
\label{fkgk}
\end{equation}
where $k_0=(|{\bf k}|^2+M^2)^{1/2}$. In the above, $M$ is the constituent
quark mass, which is calculated from dynamical chiral symetry breaking
and in general, can be momentum dependent
\cite{dnkrein,hmamspm1,hmamspm2,spmeffpot,spmbothcond}. 
Using a four point interaction for the quark operators, 
as in Nambu Jona Lasinio model, the constituent quark mass
turns out to be momentum independent \cite{amhm2004}. Also, 
a recent study \cite{dnkrein} shows the momentum dependence
of $M({\bf k})$ calculated within a color confining model, 
to be appreciable only at high momenta.
In the present work of the study of 
decay widths of the bottomonium states to open bottom mesons,
we shall assume the constituent quark mass to be momentum 
independent. 
We shall also take the approximate forms (with a small momentum
expansion) for the functions $f(|{\bf k}|)$ and $g(|{\bf k}|)$ of the 
field operator as given by 
$g(|{\bf k}|)=1/\left ({2 k_0 (k_0+M)}\right )^{1/2}
\simeq {1}/({2M}),$
and $f(|{\bf k|})=(1-g^2{\bf k}^2)^{1/2} 
\approx 1-((g^2 {\bf k}^2)/2)$ \cite{chmdwamspmwg}. 

The field operator for the constituent quark of hadron with finite 
momentum is obtained by Lorentz boosting the field 
operator of the constituent quark of hadron at rest, which 
requires the time dependence of the quark field operators.
As in the bag model, the time dependence is given by assuming 
the constituent quarks to be occupying fixed energy levels 
\cite{spm781,spm782}, so that for the $i$-th quark of a hadron
of mass $m_H$ at rest, we have
\begin{equation}
Q_i(x)=Q_i({\bf x})\exp(-i\lambda_i m_H t),
\label{thadrest}
\end{equation}
where $\lambda_i$ is the fraction of the energy (mass) of the hadron 
carried by the quark, with $\sum_i \lambda_i=1$.
For a hadron in motion with four momentum p, 
the field operators for quark annihilation and antiquark creation,
for t=0, are obtained by Lorentz boosting the field operator of the 
hadron at rest, and are given as \cite{spmdiffscat}  
\begin{equation}
Q^{(p)}({\bf x},0)=(2\pi)^{-{3}/{2}} {\int {d\bfs k S(L(p)) U({\bf k}) 
Q_I ({\bf k}+\lambda {\bf p})\exp(i({\bf k}+\lambda {\bf p}) 
\cdot{\bf x})}}
\label{qxp}
\end{equation}
and,
\begin{equation}
\tilde Q^{(p)}({\bf x},0)=(2\pi)^{-{3}/{2}} 
{\int {d\bfs k S(L(p)) V(-{\bf k}) 
\tilde Q_I (-{\bf k}+\lambda {\bf p})
\exp(-i(-{\bf k}+\lambda {\bf p}) \cdot{\bf x})}}.
\label{tldqxp}
\end{equation}
In the above, $\lambda$ is the fraction of the energy of the hadron 
at rest, carried by the constituent quark (antiquark). 
In equations (\ref{qxp}) and (\ref{tldqxp}),
$L(p)$ is the Lorentz transformation matrix, which yields 
the hadron at finite four-momentum $p$ from the hadron at rest, 
and is given as \cite{spm782}
\begin{equation}
L_{\mu 0}=L_{0 \mu}=\frac {p^\mu}{m_H};\;\;\;\;\;
L_{ij}=\delta_{ij} +\frac {p^i p^j}{m_H (p^0+m_H)},
\label{lp}
\end{equation} 
where, $\mu=0,1,2,3$ and $i=1,2,3$,
and the Lorentz boosting factor $S(L(p))$ is given as 
\begin{equation}
S(L(p))=\Bigg [\frac {(p^0+m_H)}{2m_H}\Bigg ]^{1/2}
+\Bigg [ \frac {1}{2 m_H (p^0+m_H)} \Bigg ]^{1/2} 
{\vec {\alpha}}\cdot {\vec p},
\label{slp}
\end{equation}
where, $\vec \alpha = \left (
\begin{array}{cc}  0 & \vec \sigma \\ \vec \sigma & 0
\end{array}\right)$, are the Dirac matrices.
%

\section{Partial Decay widths of the bottomonium states 
to $B\bar B$ pair in the composite model of the hadrons}

The partial decay widths of the bottomonium states,
$\Upsilon (1S)$, $ \Upsilon (2S)$, $\Upsilon (3S)$ and
$\Upsilon (4S)$ to $B\bar B$ in the hadronic matter
are studied in the present investigation.
The medium modifications of these decay widths calculated 
in the present work arise due to the medium modifications
of the decaying bottmonium state and the outgoing $B$
and $\bar B$ mesons in the hadronic medium.
In vacuum, the masses of the bottomonium states, $\Upsilon (1S)$,
$\Upsilon (2S)$, $\Upsilon (3S)$, $\Upsilon (4S)$, and the
open bottom mesons, $B$($\bar B$) are given as
9460.3 MeV, 10023.26 MeV, 10355.2 MeV, 10579.4 MeV,
and 5279 MeV respectively.
Hence, in vacuum, the lowest $\Upsilon$ state, which can decay
to $B\bar B$ is $\Upsilon (4S)$.
However, the masses of the $\Upsilon$ states as well as
$B$ and $\bar B$ mesons are modified in the hadronic medium,
due to which the partial decay widths of the 
bottomonium states to $B\bar B$ pair are  modified
in the medium. In the hadronic matter, the modification
of the $B$ meson mass turns out to be different from the
medium modification of the $\bar B$ meson mass,
due to their difference in the interactions 
with the hadronic matter. The modifications of the 
masses of the open bottom mesons arise due to 
the interactions with the nucleons, hyperons as
well as scalar mesons in the strange hadronic
matter \cite{dpambmeson}. These in-medium masses have been
calculated within an effective hadronic model,
where the chiral SU(3) model has been generalized
to SU(5) to derive the interactions of the 
$B$ and $\bar B$ mesons with the light hadrons \cite{dpambmeson}.
The bottomonium states are, on the other hand,
modified due to their interactions with the
gluon condensates in the hadronic medium.
The in-medium masses of these states 
( $\Upsilon (1S)$, $\Upsilon (2S)$, $\Upsilon (3S)$ 
and $\Upsilon (4S)$) have been calculated 
within the same effective hadronic model,
where the effect of scale symmetry breaking
of QCD through the scalar gluon condensates
are simulated by a scalar dilaton field
within the hadronic model \cite{amdpbottomonium}.
For the case of the $\Upsilon$-state at rest decaying to
$B({\bf p}) {\bar B} (-{\bf p})$, the magnitude of ${\bf p}$, 
is given by
\begin{equation}
|{\bf p}|=\Bigg (\frac{{m_\Upsilon}^2}{4}-\frac {{m_B}^2+{m_{\bar B}}^2}{2}
+\frac {({m_B}^2-{m_{\bar B}}^2)^2}{4 {m_\Upsilon}^2}\Bigg)^{1/2}.
\label{pb}
\end{equation}
In the above, the medium modifications of the masses of the bottomonium state
and the $B$ and $\bar B$ mesons are considered, so as 
to calculate the decay width of $\Upsilon \rightarrow B\bar B$
in the strange hadronic medium.

The explicit construct for the state for the bottomonium state $\Upsilon$ 
with spin projection $m$ at rest as
\begin{equation}
|\Upsilon ^{Nl} _m(\vec 0)\rangle = \int {d {\bf k}_1 
b_I^i ({\bf k}_1)^\dagger
a^{Nl}_m(\Upsilon,{\bf k}_1)\tilde {b_I}^i 
(-{\bf k}_1)|vac\rangle},
\label{upsilon}
\end{equation}
where, 
$i$ is the color index of the quark and antiquark operators.
In the present investigation, we shall assume the harmonic oscillator
wave functions for the bottomonium states and shall study
the medium modifications of the decay widths of the
bottomonium states, $\Upsilon (1S)$, $\Upsilon (2S)$,
$\Upsilon (3S)$ and $\Upsilon (4S)$, arising from
the mass modifications of the bottomonium states
as well as of the $B$ and $\bar B$ mesons.

For $\Upsilon (NS)$ (N=1,2,3,4) \cite{spmddbar80},

\begin{equation}
a^{NS}_m(\Upsilon, {\bf {k_1}})=\sigma_m u_{NS} (\bf {k_1}),
\label{am_NS}
\end{equation}

where, 
\begin{equation}
u_{1S}(\bfs k_1)
= \frac{1}{\sqrt 6} 
\Bigg (\frac {R_{\Upsilon(1S)}^2}{\pi}\Bigg )^{3/4}
\exp \left [-\frac{1}{2} R_{\Upsilon(1S)}^2 
{\bf {k_1}}^2\right ]
\label{u1s}
\end{equation}

\begin{equation}
u_{2S}(\bfs k_1)
=\frac{1}{\sqrt{6}}
{\sqrt {\frac{3}{2}}}
\Bigg({\frac{R_{\Upsilon(2S)}^2}{\pi}}\Bigg)^{3/4}
\left(\frac{2}{3}R_{\Upsilon(2S)}^2\bfs k_1^2-1\right)
\exp\left[-\frac{1}{2}R_{\Upsilon(2S)}^2\bfs k_1^2\right].
\label{u2s}
\end{equation}

\begin{equation}
u_{3S}(\bfs k_1)
=\frac{1}{\sqrt{6}}
{\sqrt {\frac{15}{8}}}
\Bigg({\frac{R_{\Upsilon(3S)}^2}{\pi}}\Bigg)^{3/4}
\left(1-\frac{4}{3}R_{\Upsilon(3S)}^2\bfs k_1^2
+\frac{4}{15}R_{\Upsilon(3S)}^4\bfs k_1^4
\right)
\exp\left[-\frac{1}{2}R_{\Upsilon(3S)}^2\bfs k_1^2\right].
\label{u3s}
\end{equation}

\begin{eqnarray}
u_{4S}(\bfs k_1)
&=&-\frac{1}{\sqrt{6}}
{\frac{ \sqrt {35}}{4}}
\Bigg({\frac{R_{\Upsilon(4S)}^2}{\pi}}\Bigg)^{3/4}
\left(1-2 R_{\Upsilon(4S)}^2\bfs k_1^2
+\frac{4}{5}R_{\Upsilon(4S)}^4\bfs k_1^4
-\frac{8}{105}R_{\Upsilon(4S)}^6\bfs k_1^6
\right)\nonumber \\
&\times &
\exp\left[-\frac{1}{2}R_{\Upsilon(4S)}^2\bfs k_1^2\right].
\label{u4s}
\end{eqnarray}
In the above, the factor $\frac {1}{\sqrt 6}$ refers to normalization
factor arising from degeneracy factors due to color (3) and 
spin (2)  of the quarks and antiquarks.

The $B^0$ and $\bar {B^0}$ states, with finite momenta, 
are explicitly given as
\begin{equation}
|B^0 ({\bf p}')\rangle = { \int {d_I^{i_2}({\bf k}_3
+\lambda_1 {\bf p}')
^\dagger u_{B}({\bf k}_3)\tilde {b_I}^{i_2} 
(-{\bf k}_3 +\lambda_2 {\bf p}') d\bfs k_3}}.
\label{b0}
\end{equation}
and 
\begin{equation}
|\bar {B^0} ({\bf p})\rangle 
={ \int {b_I^{i_1}}({\bf k}_2+\lambda_2 {\bf p})
^\dagger u_{B}({\bf k}_2)\tilde {d_I}^{i_1} 
(-{\bf k}_2 +\lambda_1 {\bf p}) d\bfs k_2}
\label{b0bar}
\end{equation}
In the above, 
\begin{equation}
u_{B}({\bf k})=\frac{1}{\sqrt{6}}\Bigg 
(\frac {{R_B}^2}{\pi} \Bigg)^{3/4}
\exp\Bigg(-\frac {{R_B}^2 {{\bf k}}^2}{2}\Bigg).
\label{ub}
\end{equation}
To calculate the partial decay widths of the decay process
$\Upsilon \rightarrow B\bar B$, we need to know the
values of $\lambda_1$ and $\lambda_2$, the fractions of 
energy of the hadron carried by its constituent quark and antiquark.
These are calculated by assuming that the binding energy of the hadron 
as shared by the quark/antquark are {\it inversely} 
proportional to the quark/antiquark masses
\cite{spm782,chmdwamspmwg}. The energies of
$d(\bar d)$ and $\bar b (b)$ in ${\bar B} (B)$ meson are 
then given as
\cite{spm782},
\begin{equation}
\omega_1=M_d+\frac{M_b}{M_b+M_d}\cdot(m_B-M_b-M_d)
\label{omega1}
\end{equation}
and,
\begin{equation}
\omega_2=M_b+\frac{M_d}{M_b+M_d}\cdot(m_B-M_b-M_d),
\label{omega2}
\end{equation}
with 
\begin{equation}
\lambda_i=\frac{\omega_i}{m_B}.
\label{lambda12}
\end{equation} 

The motivation for the assumption that the 
contributions from the quark (antiquark) to the binding 
energy of the hadron to be inversely proportional
to the mass of the quark (antiquark) as in equations (\ref{omega1})
and (\ref{omega2}) is as follows. In fact, in general,
the contributions to the binding energy of the bound state composed
of particles of 1 and 2, with masses $m_1$ and $m_2$, are assumed to be
given as ${\mu}/{m_i}$, $i=1,2$, multiplied by the binding energy 
of the bound state, where, $\mu$ is the reduced mass
of the system, calculated from ${1}/{\mu}={1}/{m_1} +
{1}/{m_2}$.
In other words, the contributions from the particles
to binding energy are inversely proportional to their masses,
and the total binding energy is the sum of the individual 
contributions, i.e., $BE=\Big (({\mu}/{m_1})+({\mu}/{m_2})
\Big )\times BE =BE$, as it should be.
The reason for making this assumption comes from the example of
hydrogen atom, which is the bound state of the proton and the electron.
As the mass of proton is much larger as compared to the mass of the 
electron, the binding energy contribution from the electron is
$\frac{\mu}{m_e}\times BE \simeq BE$ of hydrogen atom, and the
contribution from the proton is $\frac{\mu}{m_p}\times BE$, 
which is negligible as compared to the total 
binding energy of hydrogen atom, since $m_p >> m_e$. 
With this assumption, the binding 
energies of the heavy-light mesons, e.g., $D$ and $(\bar D)$ mesons 
\cite{chmdwamspmwg}, as well as for  $B$ and $(\bar B)$ mesons,
mostly arise from the contribution from the light quark (antiquark).

We next evaluate the matrix element of the quark-antiquark pair creation
part, $\int {\cal H}_{Q^\dagger\tilde Q}({\bf x},t=0)d{\bf x}$,
of the Dirac Hamiltonian density,
between the initial and the final states for the reaction
$\Upsilon \rightarrow {\bar B}^0 ({\bf p})+{B^0}({\bf p}')$.
The Dirac Hamiltonian density is given as 
\begin{equation}
{\cal H}=\psi (x) ^\dagger (-i{\vec \alpha} \cdot {\vec \bigtriangledown} 
+\beta m_Q)\psi (x),
\end{equation}
where $\vec \alpha$ and $\beta$ are the Dirac matrices, with $\vec \alpha$
defined following equation (\ref{slp}) and $\beta = diag (I,-I)$.
In the above, $\psi (x)$
is the field operator for the constituent quark, $Q$ with mass $m_Q$,
which is given by equation (\ref{qx}) for t=0.
The relevant part of the quark pair creation term for the decay
process $\Upsilon \rightarrow {\bar B}^0 ({\bf p})+{B^0}({\bf p}')$
 is through the $d \bar d$ creation.
From equations (\ref{qxp}) and (\ref{tldqxp})
we can write down
${\cal H}_{d^\dagger\tilde d}({\bf x},t=0)$, and then
integrate over ${\bf x}$ to obtain the expression  
\begin{eqnarray}
&& {\int {{\cal H}_{d^\dagger\tilde d}({\bf x},t=0)d{\bf x}}}
\nonumber\\
&=&
{\int {d\bfs k d\bfs k'd_I^i({\bf k}+\lambda_1{\bf p}')^\dagger 
U({\bf k})^\dagger S(L(p'))^\dagger
\delta(-{\bf k}'+\lambda_1{\bf p}+{\bf k}+\lambda_1{\bf p}')}}
\nonumber\\
&\times & (\bfm \alpha \cdot ({\bf k}+\lambda_1{\bf p}')+\beta M_d)
S(L(p))V(-{\bf k}')\tilde d_I^i(-{\bf k}'+\lambda_1{\bf p}),
\label{hint}
\end{eqnarray}
where, $S(L(p))$ and $S(L(p'))$ in the above equation, 
correspond to the hadrons
with finite momenta, $p$ and $p'$, 
i.e., $\bar {B^0}$ and $B^0$ mesons,
and $M_d$ is the constituent mass of the $d$ quark.
$S(L(p))$
has already been defined in equation (\ref{slp}), 
and, $U({\bf k})$ and $V({\bf k})$ are given by 
equation (\ref{ukvk}).

\noindent From equation (\ref{hint}) we can then evaluate that
\begin{equation}
\langle {\bar {B^0}} ({\bf p}) | \langle B^0 ({\bf p}')|
{ \int {{\cal H}_{d^\dagger\tilde d}({\bf x},t=0)d{\bf x}}}
|{\Upsilon(NS) }_m (\vec 0) \rangle
=\delta({\bf p}+{\bf p}'){\int {d\bfs k_1 
A^{\Upsilon(NS)}_m({\bf p},{\bf k}_1)}},
\label{upslnbbar}
\end{equation}
using the explicit forms of the $\Upsilon$-states
and $\bar {B^0}$ and $B^0$ states. 
We obtain the form of $A^{\Upsilon}_m({\bf p},{\bf k}_1)$, 
including summing over color,
\begin{eqnarray}
 A^{\Upsilon(NS)}_m({\bf p},{\bf k}_1)
&=& 3 u_{\bar {B^0}}({\bf k}) u_{B^0}({\bf k})
\cdot {\rm {Tr}} \big [a_m(\Upsilon(NS),{\bf k}_1)
U({\bf k})^\dagger  \nonumber\\
&\times& S(L(p'))^\dagger
(\bfm\alpha\cdot{\bf {\tilde q}}+\beta M_d)
S(L(p))V(-{\bf k})
\big ],
\label{ampk}
\end{eqnarray}
where, ${\bfm k} = {\bf k}_1 -\lambda_2{\bf p}$,
${\bfm {\tilde q}} = {\bf k}_1 -{\bf p}$ and
$\bfm p'=-\bfm p$.

We shall now simplify $A^{\Upsilon(NS)}_m({\bf p},{\bf k}_1)$. 
Firstly, since the 
$B(\bar B)$ mesons are completely nonrelativistic, we shall be assuming 
that $S(L(p))$ and $S(L(p'))$ as identity. 
The integral in the R.H.S. of the equation 
(\ref{upslnbbar})
can be written as,
\begin{eqnarray}
\int d {\bfm {k_1}} A^{\Upsilon(NS)}_m({\bf p},{\bf k}_1)
=3\int d {\bfm {k_1}} u_{\bar {B^0}}({\bf k}) u_{B^0}({\bf k})
\cdot {\rm {Tr}} \big [a_m(\Upsilon(NS),{\bf k}_1)
B({\bfm k}, {\bfm {\tilde q}})
\big ],
\label{ampk1}
\end{eqnarray}
where,
\begin{equation}
B(\bfm k, {\bfm {\tilde q}})={\bfm \sigma}\cdot {\bfm {\tilde q}}
-\Big ( 2 (\bfm k \cdot {\bfm {\tilde q}})g^2 +f(\bfm k) \Big )
{\bfm \sigma}\cdot {\bfm k}.
\end{equation}
We use the approximate forms of $f$ and $g$ at small momentum, i.e., 
$f\approx 1-\frac {g^2 {\bf k}^2}{2}$, and $2M_d g \approx 1$,
for simplifying the integral
given by equation (\ref{ampk1}). After simplification, this 
integral can be written as
\begin{equation}
\int d {\bfm {k_1}} A^{\Upsilon(NS)}_m({\bf p},{\bf k}_1)
=6c_{\Upsilon (NS)} \exp\big [(a_{\Upsilon (NS)} 
b_{\Upsilon (NS)}^2 -{R_B}^2 
{\lambda_2}^2)|{\bfm p}|^2\big]
\int d {\bfm {k_1}} T^{\Upsilon(NS)}_m({\bf p},{\bf k}_1),
\end{equation}
where, 
$T^{\Upsilon(NS)} ({\bf p},{\bf k}_1)$, for $N=1,2,3,4$, are given as,
\begin {eqnarray}
T^{\Upsilon(1S)}_m({\bf p},{\bf k}_1) 
&=&\frac{1}{2} {\rm {Tr}} \big [\sigma_m
B({\bfm k}, {\bfm {\tilde q}})\big],
\nonumber \\
T^{\Upsilon(2S)}_m({\bf p},{\bf k}_1) 
&=&\frac{1}{2} {\rm {Tr}} \big [\sigma_m
B({\bfm k}, {\bfm {\tilde q}})\big]
\cdot \left(\frac{2}{3}R_{\Upsilon(2S)}^2\bfs k_1^2-1\right),
\nonumber \\
T^{\Upsilon(3S)}_m({\bf p},{\bf k}_1) 
&=&\frac{1}{2} {\rm {Tr}} \big [\sigma_m
B({\bfm k}, {\bfm {\tilde q}})\big]
\cdot
\left(1-\frac{4}{3}R_{\Upsilon(3S)}^2\bfs k_1^2
+\frac{4}{15}R_{\Upsilon(3S)}^4\bfs k_1^4
\right),
\nonumber \\
T^{\Upsilon(4S)}_m({\bf p},{\bf k}_1) 
&=&\frac{1}{2} {\rm {Tr}} \big [\sigma_m
B({\bfm k}, {\bfm {\tilde q}})\big]
\cdot
\left(1-2 R_{\Upsilon(4S)}^2\bfs k_1^2
+\frac{4}{5}R_{\Upsilon(4S)}^4\bfs k_1^4
-\frac{8}{105}R_{\Upsilon(4S)}^6\bfs k_1^6
\right).
\label{tmupsln}
\end{eqnarray}

\noindent In the above, the parameters $a_{\Upsilon (NS)}$ and 
$b_{\Upsilon (NS)}$
are given as
\begin{equation}
a_{\Upsilon (NS)}=\frac{1}{2}R_{\Upsilon (NS)}^2+R_B^2; \;\;\;\; 
b_{\Upsilon (NS)}=R_B^2\lambda_2/a_{\Upsilon (NS)},
\label{abupsln}
\end{equation}
with $R_{\Upsilon (NS)}$ as the radius of the bottomonium state,
$\Upsilon (NS)$ $(N=1,2,3,4)$, and,
\begin{eqnarray}
c_{\Upsilon (1S)}&=&\frac{1}{6\sqrt{6}}\cdot
\left(\frac{R_{\Upsilon(1S)}^2}{\pi}\right)^{{3}/{4}}
\cdot\left(\frac{R_B^2}{\pi}\right)
^{{3}/{2}},
\nonumber \\
c_{\Upsilon (2S)}&=&\frac{1}{6\sqrt{6}}
\sqrt {\frac{3}{2}}
\left(\frac{R_{\Upsilon(2S)}^2}{\pi}\right)^{{3}/{4}}
\cdot \left(\frac{R_B^2}{\pi}\right)^{{3}/{2}},
\nonumber \\
c_{\Upsilon (3S)}&=&\frac{1}{6\sqrt{6}}
\sqrt {\frac{15}{8}}
\left(\frac{R_{\Upsilon(3S)}^2}{\pi}\right)^{{3}/{4}}
\cdot\left(\frac{R_B^2}{\pi}\right)^{{3}/{2}},
%
\nonumber \\
c_{\Upsilon (4S)}&=&\frac{1}{6\sqrt{6}}
\left(\frac{\sqrt {35}}{4}\right)
\left(\frac{R_{\Upsilon(4S)}^2}{\pi}\right)^{{3}/{4}}
\cdot\left(\frac{R_B^2}{\pi}\right)^{{3}/{2}}.
\end{eqnarray}
We now change the integration variable to $\bfs q$ 
in equation (\ref{ampk1}) with the
substitution $\bfs k_1=\bfs q+{b_{\Upsilon(NS)}}\bfs p$ and write
\begin{equation}
{\int {A^{\Upsilon(NS)}_m({\bf p},{\bf k}_1)d\bfs k_1}}
=6c_{\Upsilon(NS)}\exp[(a_{\Upsilon (NS)} b_{\Upsilon(NS)}^2
-\lambda_2^2 R_B^2)
|{\bf p}|^2]\cdot 
{\int {\exp(-{a_{\Upsilon(NS)}}\bfs q^2)T_m^{\Upsilon(NS)}d\bfs q}},
\label{ampqt}
\end{equation}
where, $T_m^{\Upsilon(NS)}$ in the above equation,
is the expression $T^{\Upsilon(NS)}_m({\bf p},{\bf k}_1)$
given by equation (\ref{tmupsln}), rewritten in terms 
of $\bf q$.
We next proceed to simplify the above integral,
by using the fact that the terms odd in $\bfs q$ in equation 
(\ref{ampqt}) will vanish. Also, 
using  
${\bf q}_i{\bf q}_jG(|{\bf q}|)\equiv \frac{1}{3}\delta_{ij}{\bf q}^2
G(|{\bf q}|)$ and 
${\bf q}_i{\bf q}_j  {\bf q}_k{\bf q}_m  
G(|{\bf q}|)\equiv \frac{1}{15}(\delta_{ij}\delta _{km}
+\delta_{ik}\delta _{jm}+\delta_{im}\delta _{jk})
{\bf q}^4 G(|{\bf q}|)$, where, $G(|{\bf q}|)$ is an even
function of $\bf q$, 
$T_m^{\Upsilon(NS)}(\bfs p,\bfs q)$ 
in the above integrand can be recast into the form 
\begin{eqnarray}
T_m^{\Upsilon (NS)}(\bfs p,\bfs q)
& \equiv & p^m \big [F_0^{\Upsilon (NS)}(|\bfs p|)
+F_1^{\Upsilon (NS)}(|\bfs p|)|\bfs q|^2
+F_2^{\Upsilon (NS)}(|\bfs p|)(|\bfs q|^2)^2
\nonumber \\
&+& F_3^{\Upsilon (NS)}(|\bfs p|)(|\bfs q|^2)^3
+F_4^{\Upsilon (NS)}(|\bfs p|)(|\bfs q|^2)^4
\big ]
\label{tmupslnNS}
\end{eqnarray}
The coefficients, $F_i^{\Upsilon (NS)} (i=0,1,2,3,4, {\rm and}
N=1,2,3,4)$
are given as
\begin{eqnarray}
F^{\Upsilon (1S)}_0 &= &(\lambda_2-1)
\nonumber \\
&-& 2g^2 |{\bfs p}|^2(b_{\Upsilon(1S)}-\lambda_2)
\left(\frac{3}{4}b_{\Upsilon (1S)}^2
-(1+\frac{1}{2}\lambda_2)b_{\Upsilon (1S)}
+\lambda_2-\frac{1}{4}\lambda_2^2\right),
\nonumber\\
F^{\Upsilon (1S)}_1&=&g^2\left[-\frac{5}{2}b_{\Upsilon (1S)}
+\frac{2}{3}
+\frac{11}{6}\lambda_2\right],
\nonumber\\
F^{\Upsilon (1S)}_2&=&0,
\nonumber \\
F^{\Upsilon (1S)}_3&=&0,
\nonumber \\
F^{\Upsilon (1S)}_4&=&0,
\label{fiupsln1s}
\end{eqnarray}
\begin{eqnarray}
F^{\Upsilon (2S)}_0
&=&\left(\frac{2}{3}R_{\Upsilon (2S)}^2
b_{\Upsilon (2S)}^2|{\bfs p}|^2-1\right)
F^{\Upsilon (1S)}_0,
\nonumber\\
F^{\Upsilon (2S)}_1 &=& \frac{2}{3}R_{\Upsilon (2S)}^2
 F^{\Upsilon (1S)} _0
+\left(\frac{2}{3}R_{\Upsilon (2S)}^2 b_{\Upsilon (2S)}^2
|{\bfs p}|^2
-1\right)
F^{\Upsilon (1S)}_1
\nonumber \\ 
&-& \frac{8}{9}R_{\Upsilon (2S)}^2
{b_{\Upsilon (2S)}}g^2 |{\bfs p}|^2\left[\frac{9}{4}b_{\Upsilon (2S)}^2
-b_{\Upsilon (2S)}\left(2+\frac{5}{2}\lambda_2
\right)+2\lambda_2+\frac{1}{4}\lambda_2^2\right],\nonumber\\
F^{\Upsilon (2S)}_2 &= & \frac{2}{3}R_{\Upsilon (2S)}^2g^2
\left[-\frac{7}{2}b_{\Upsilon (2S)}
+\frac{2}{3}+\frac{11}{6}
\lambda_2\right],
\nonumber\\
F^{\Upsilon (2S)}_3 &= & 0,
\nonumber\\
F^{\Upsilon (2S)}_4 &=& 0,
\label{fiupsln2s}
\end{eqnarray}

\begin{eqnarray}
 F^{\Upsilon (3S)}_0
&=&\left(    
-1+\lambda_2 + \frac {g^2 |{\bf p}|^2}{2} 
(b_{\Upsilon (3S)}-\lambda_2)^2 
(3 b_{\Upsilon (3S)} +\lambda_2 -4) \right )
\nonumber \\
& \times &
\Bigg (1-\frac{4}{3} R_{\Upsilon (3S)}^2 
b_{\Upsilon (3S)}^2 |{\bf p}|^2
+\frac{4}{15} R_{\Upsilon (3S)}^4 
b_{\Upsilon (3S)}^4 {|\bf p|}^4\Bigg),
\nonumber\\
F^{\Upsilon (3S)}_1
&=&\frac {4}{3} R_{\Upsilon (3S)}^2 (1-\lambda_2)
\Big ( 1-\frac {2}{3} b_{\Upsilon (3S)}^2 R_{\Upsilon (3S)}^2
|{\bf p}|^2 \Big ) 
+\frac {g^2}{6} (3b_{\Upsilon (3S)}-7\lambda_2 +4)
\nonumber \\
& +&\frac {g^2 {|\bf p|}^2  R_{\Upsilon (3S)}^2}{9}
\Bigg [ 
(-3b_{\Upsilon (3S)} +7\lambda_2 -4 )
b_{\Upsilon (3S)}^2
\nonumber \\ 
&+& 
4 (3b_{\Upsilon (3S)} -\lambda_2 -2 )
(b_{\Upsilon (3S)}-\lambda_2) 
(-2 b_{\Upsilon (3S)} + 3 \lambda_2)
\Bigg ]
\nonumber \\
&+&\frac {2 g^2 {|\bf p|}^4 R_{\Upsilon (3S)} ^4 
b_{\Upsilon (3S)}^2}{45}
\Bigg [ (3b_{\Upsilon (3S)} -7\lambda_2 +4 )
b_{\Upsilon (3S)}^2
\nonumber\\
&+& 
4 (3b_{\Upsilon (3S)}-4\lambda_2) 
(3b_{\Upsilon (3S)} -\lambda_2 -2 )
( b_{\Upsilon (3S)} - \lambda_2) 
\Bigg ]
\nonumber \\
 F^{\Upsilon (3S)}_2
&=& \frac {4}{15} (\lambda_2-1) R_{\Upsilon (3S)}^4 
-\frac {2}{9} g^2 R_{\Upsilon (3S)}^2 
(9b_{\Upsilon (3S)}-7\lambda_2 +4) 
\nonumber \\
&+& \frac {g^2 R_{\Upsilon (3S)}^4 |{\bf p}|^2}{15} 
\Bigg [ 8 b_{\Upsilon (3S)}^3 -\frac {8}{3} 
b_{\Upsilon (3S)} (b_{\Upsilon (3S)}-\lambda_2) 
(3b_{\Upsilon (3S)} +\lambda_2 -4) 
\nonumber \\
&+& 2 (b_{\Upsilon (3S)}-\lambda_2) ^2 
(3 b_{\Upsilon (3S)} +\lambda_2 -4)
+ 14 b_{\Upsilon (3S)} ^2 
(b_{\Upsilon (3S)}-\lambda_2) 
\nonumber \\
&-& 
\frac {88}{15} b_{\Upsilon (3S)}^2 
(3 b_{\Upsilon (3S)} -\lambda_2 -2)
\Bigg ],
\nonumber \\
 F^{\Upsilon (3S)}_3
&=&\frac {2 g^2}{45} R_{\Upsilon (3S)} ^4 
(15 b_{\Upsilon (3S)} -7 \lambda_2 +4),
\nonumber \\
F^{\Upsilon (3S)}_4 &=& 0,
\label{fiupsln3s}
\end{eqnarray}
and, 
\begin{eqnarray}
F^{\Upsilon (4S)}_0
&=& \frac {1}{2} (b_{\Upsilon (4S)}-1) (b_{\Upsilon (4S)}-\lambda_2)
(3 b_{\Upsilon (4S)} +\lambda_2 -4) g^2 |{\bf p}|^2
\nonumber \\
&&\times  
\Bigg (1-2 R_{\Upsilon (4S)}^2 
b_{\Upsilon (4S)}^2 |{\bf p}|^2
+\frac{4}{5} R_{\Upsilon (4S)}^4 
b_{\Upsilon (4S)}^4 {|\bf p|}^4
-\frac{8}{105} R_{\Upsilon (4S)}^6 
b_{\Upsilon (4S)}^6 {|\bf p|}^6
\Bigg)
\nonumber\\
F^{\Upsilon (4S)}_1
&=&
\frac {g^2}{6} \Big ( 9 (b_{\Upsilon (4S)}-1)
- 2 (3 b_{\Upsilon (4S)} -\lambda_2 -2) \Big )
\nonumber \\
&+& \frac {g^2 {|\bf p|}^2  R_{\Upsilon (4S)}^2}{3}
\Bigg [ (-5 b_{\Upsilon (4S)} +3) (3b_{\Upsilon (4S)} 
+\lambda_2 -4 )(b_{\Upsilon (4S)}-\lambda_2) 
\nonumber \\
&-& 
9 b_{\Upsilon (4S)} ^2 (b_{\Upsilon (4S)}-1) 
+ 2 b_{\Upsilon (4S)} (3 b_{\Upsilon (4S)} -\lambda_2 -2) 
( 3 b_{\Upsilon (4S)} -2)
\Bigg ]
\nonumber \\
&+&\frac {4 g^2 {|\bf p|}^4 R_{\Upsilon (4S)} ^4 b_{\Upsilon (4S)}^2}{15}
\Bigg [ (7b_{\Upsilon (4S)} -5) (3b_{\Upsilon (4S)} 
+\lambda_2 -4 )(b_{\Upsilon (4S)}-\lambda_2) 
\nonumber \\
&+& \frac {9}{2} (b_{\Upsilon (4S)} -1) b_{\Upsilon (4S)} ^2 
-  b_{\Upsilon (4S)} ( 5 b_{\Upsilon (4S)} - 4)
(3b_{\Upsilon (4S)} -\lambda_2 -2 )\Bigg ]
\nonumber \\
&-& 
\frac {8 g^2 {|\bf p|}^6 R_{\Upsilon (4S)} ^6 b_{\Upsilon (4S)}^4}{105}
\Bigg [ \frac {1}{2}(9b_{\Upsilon (4S)} -7) 
(3b_{\Upsilon (4S)} +\lambda_2 -4 )(b_{\Upsilon (4S)}-\lambda_2) 
\nonumber \\
&+&\frac {3}{2} b_{\Upsilon (4S)} ^2 (b_{\Upsilon (4S)}-1) 
-\frac {1}{3} b_{\Upsilon (4S)} (3 b_{\Upsilon (4S)} 
-\lambda_2 -2) (7b_{\Upsilon (4S)} -6)
\Bigg ]
\nonumber\\
 F^{\Upsilon (4S)}_2
&=&\frac {1}{3} g^2 R_{\Upsilon (4S)}^2 (-9 b_{\Upsilon (4S)} 
-2 \lambda_2 +5)
\nonumber \\
&+& \frac {4}{5} g^2 R_{\Upsilon (4S)} ^4 |{\bf p}|^2
\Bigg [ b_{\Upsilon (4S)} ^2 (7 b_{\Upsilon (4S)} -5) 
\nonumber \\
&+& \frac {1}{6} (3 b_{\Upsilon (4S)} +\lambda_2 -4)
(b_{\Upsilon (4S)} -\lambda_2) (7 b_{\Upsilon (4S)} -3) 
\nonumber \\
 &-& \frac {2}{15}  b_{\Upsilon (4S)} 
(3 b_{\Upsilon (4S)} -\lambda_2 -2)
(21 b_{\Upsilon (4S)} -10) \Bigg ]
\nonumber \\
&+& \frac {4}{5} g^2 R_{\Upsilon (4S)} ^6 |{\bf p}|^4 b_{\Upsilon (4S)} ^ 2
\Bigg [ -\frac {1}{7} b_{\Upsilon (4S)}^2 (9 b_{\Upsilon (4S)} -7)
\nonumber \\
&-&\frac {4}{15} b_{\Upsilon (4S)} (b_{\Upsilon (4S)} -\lambda_2 ) 
(3 b_{\Upsilon (4S)} +\lambda_2 -4)
\nonumber \\
&& -\frac {1}{3} (b_{\Upsilon (4S)}-1) (b_{\Upsilon (4S)} -\lambda_2) 
(3 b_{\Upsilon (4S)} +\lambda_2 -4) 
\nonumber \\
&+ & \frac {2}{105} b_{\Upsilon (4S)} (3b_{\Upsilon (4S)} -\lambda_2 -2) 
(45 b_{\Upsilon (4S)} -28 ) \Bigg ],  
\nonumber \\
 F^{\Upsilon (4S)}_3
&=&\frac {2 g^2}{15} R_{\Upsilon (4S)} ^4 (15 b_{\Upsilon (4S)} 
+2 \lambda_2 -5)
\nonumber \\
&+&\frac {4}{5} g^2 R_{\Upsilon (4S)} ^6 |{\bf p}|^2
\Bigg [ -\frac {4}{5} b_{\Upsilon (4S)}^3 
- (b_{\Upsilon (4S)}-1) b_{\Upsilon (4S)} ^2 
\nonumber \\
&-&\frac {2}{21} b_{\Upsilon (4S)} 
(b_{\Upsilon (4S)}-\lambda_2) (3 b_{\Upsilon (4S)} +\lambda_2 -4)
\nonumber \\
&-&\frac {1}{21} (b_{\Upsilon (4S)}-1) (b_{\Upsilon (4S)}-\lambda_2) 
(3 b_{\Upsilon (4S)} +\lambda_2 -4)) 
\nonumber \\
&+& \frac {2}{105} b_{\Upsilon (4S)} (3 b_{\Upsilon (4S)} -\lambda_2 -2)
(27b_{\Upsilon (4S)} -10) \Bigg ] ,
\nonumber \\
 F^{\Upsilon (4S)}_4
&=&-\frac {4 g^2 R_{\Upsilon (4S)} ^6 }{35\times 9} 
(21b_{\Upsilon (4S)} + 2 \lambda_2 -5).
\label{fiupsln4s}
\end{eqnarray}
\noindent On performing the integration over $\bfs q$, 
equation (\ref{ampqt})  yields
\begin{equation}
\int A^{\Upsilon (NS)}_m({\bf p},{\bf k}_1)d {\bf k}_1
 =A^{\Upsilon (NS)}(|{\bf p}|){\bf p}_m,
\label{ap3}
\end{equation}
where, $A^{\Upsilon (NS)}(|\bfs p|)$ is given as
\begin{eqnarray}
A^{\Upsilon (NS)}(|{\bf p}|) &= & 
6c_{\Upsilon (NS)}\exp[(a_{\Upsilon (NS)} 
{b^2_{\Upsilon (NS)}}
-R_B^2\lambda_2^2)|{\bf p}|^2]
\cdot\Bigg(\frac{\pi}{a_{\Upsilon (NS)}}\Bigg)^{{3}/{2}}
\nonumber \\
&\times &
\Bigg[F_0^{\Upsilon (NS)}+\frac{3}{2a_{\Upsilon (NS)}}
\cdot F_1^{\Upsilon (NS)}
+ \frac{15}{4a_{\Upsilon (NS)}^2}\cdot F_2^{\Upsilon (NS)}
\nonumber \\
&+&
\frac{105}{8a_{\Upsilon (NS)}^3} \cdot F_3^{\Upsilon (NS)}
+ \frac{105 \times 9}{16a_{\Upsilon (NS)}^4}
\cdot F_4^{\Upsilon (NS)}
\Bigg].
\label{ap}
\end{eqnarray}
With $<f|S|i>=\delta_4(P_f-P_i)M_{fi}$ 
we then have for bottomonium state, $\Upsilon (NS)$ of spin $m$,
\begin{equation}
M_{fi}=2\pi\cdot(-iA^{\Upsilon (NS)}(|{\bf p}|)){\bf p}_m.
\label{mfi}
\end{equation}
In the present work, we shall be studying the in-medium decay widths
of the bottomonium state, $\Upsilon (NS)$ to $B\bar B$ pair,
arising due to the mass modifications of the bottomonium 
state and the $B$ and $\bar B$ states.

The expression obtained for the partial decay width 
of the bottomonium state
decaying at rest to $B^0{\bar {B^0}}$ pair, after averaging
over spin, is given as \cite{chmdwamspmwg}, 
\begin{eqnarray}
\Gamma(\Upsilon (NS) \rightarrow B^0 {\bar {B^0}})
&=& \gamma_\Upsilon ^2 \frac{1}{2\pi} 
\int \delta(m_{\Upsilon (NS)}-p^0_{B^0}-p^0_{\bar {B^0}})
{|M_{fi}|^2}_{av}
\cdot 4\pi |{\bfs p}_{B^0}|^2 d|{\bfs p}_{\bar {B^0}}| 
\nonumber\\
&=& \gamma_\Upsilon ^2\frac{8\pi^2}{3}|{\bf p}|^3
\frac {p^0_{B^0} p^0_{\bar {B^0}}}{m_{\Upsilon (NS)}}
A^{\Upsilon (NS)}(|{\bf p}|)^2
\label{gammaupslnb0b0b}
\end{eqnarray}
In the above, $p^0_{B^0}=\big(m_{B^0}^2+|{\bf p}|^2\big)^{{1}/{2}}$, 
$p^0_{\bar {B^0}}=\big(m_{\bar {B^0}}^2+|{\bf p}|^2\big)^{{1}/{2}}$, 
and, $|\bfs p|$ is the magnitude of the momentum of the outgoing 
$B^0 (\bar {B^0})$ mesons. The decay of 
$\Upsilon (NS)$ to $B^+ B^-$ proceeds through a 
$u \bar u$ pair creation and the decay width 
(\ref{gammaupslnb0b0b}) is modified to 
\begin{eqnarray}
\Gamma(\Upsilon (NS) \rightarrow B^+ B^-)
=&& \gamma_\Upsilon^2\frac{8\pi^2}{3}\cdot|{\bf p}|^3
\frac {{p^0}_{B^+} {p^0}_{B^-}}{m_{\Upsilon (NS)}}
A^{\Upsilon (NS)}(|{\bf p}|)^2
\label{gammaupslnbpbm}
\end{eqnarray}
In the above, $p^0_{B^\pm}=\big(m_{B^\pm}^2+|{\bf p}|^2\big)^{{1}/{2}}$, 
and, $|\bfs p|$ is the magnitude of the momentum of the outgoing 
$B^\pm$ mesons.
The parameter $\gamma_\Upsilon$, has been introduced in the
expressions for the decay widths of $\Upsilon (NS) \rightarrow
B^0 \bar {B^0} (B^+ B^-)$, which is a measure of
the production strength of the $B\bar B$ pair from
the $\Upsilon$-state through light quark antiquark pair 
($d\bar d$ or $u\bar u$) creation. To study the
decay width of quarkonia using a light quark pair creation model, 
namely, $^3P_0$ model, such a pair creation strength parameter, 
$\gamma$ has been introduced in Ref. \cite{barnes,friman}, 
which was fitted to the observed decay width of the meson.
In the present investigation of the bottomonium 
decay widths, the parameter, $\gamma_\Upsilon$
is fitted from the vacuum decay width for the
channel $\Upsilon (4S) \rightarrow B\bar B$
($\Upsilon (4S)$ is the lowest $\Upsilon$-state
which decays to $B\bar B$ in vacuum).
In the present work, we study the effects
of the medium effects on the decay widths
of the $\Upsilon \rightarrow B\bar B$,
arising from the medium modifications
of the masses of $\Upsilon$ as well as the $B$ and  
$\bar B$ mesons, and explore the possibility 
whether the decays of the lower (excited) states of bottomonium
can become kinematically possible with the medium
effects.


\begin{figure}
\vskip -0.9in
\includegraphics[width=18cm,height=17cm]{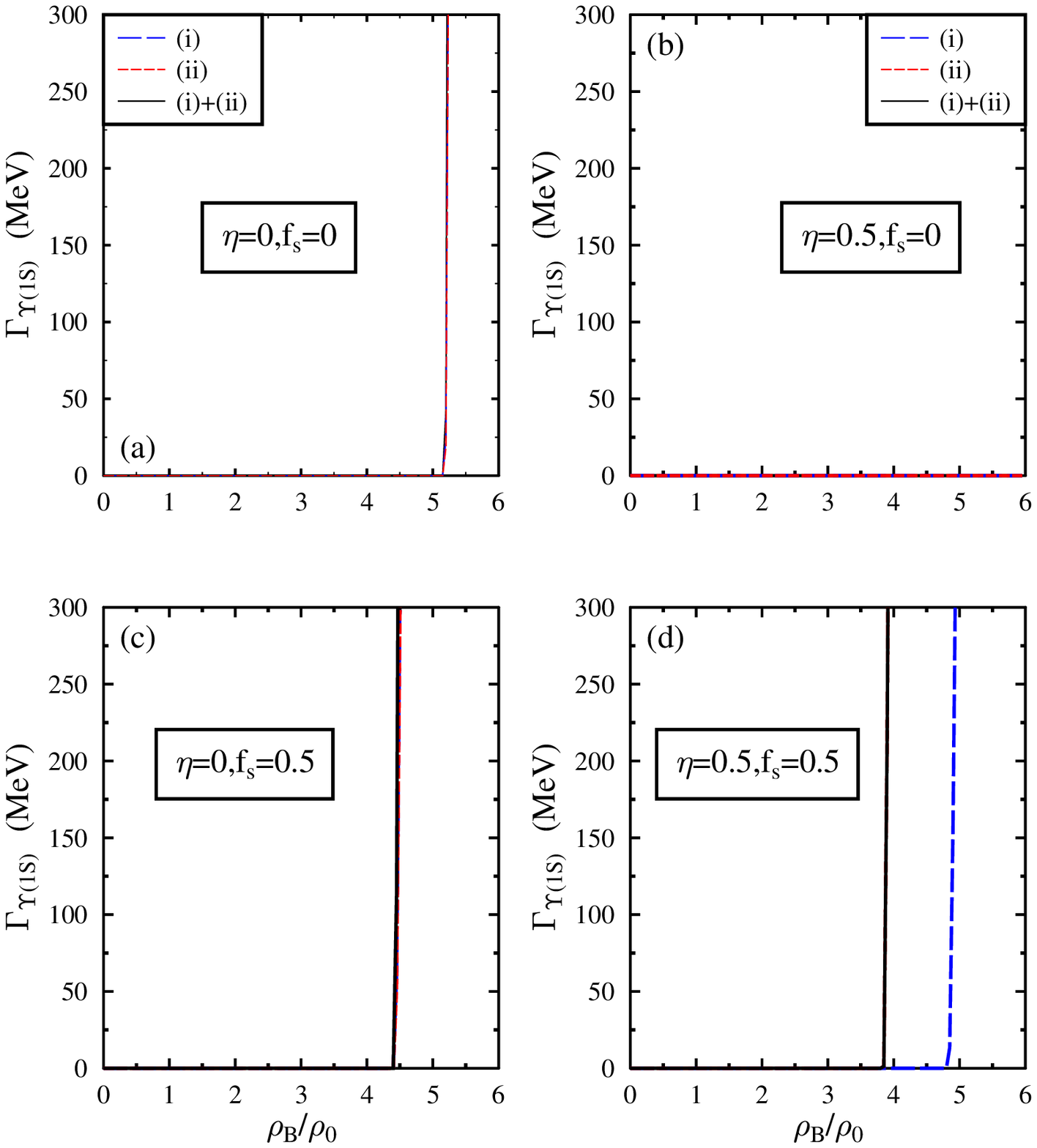}
\vskip -0.9in
\caption{(Color online) The partial decay widths of $\Upsilon (1S)$
calculated using the present model for composite hadrons,
to (i) $B^+B^-$, (ii) $B^0\bar {\rm B^0}$ and (iii) the sum of the 
two channels ((i)+(ii)) in the isospin symmetric (asymmetric)
nuclear matter and strange hadronic matter. In (a) and (c), the 
decay widths are shown for isospin symmetric ($\eta$=0) matter,
where the channels (i) and (ii) are observed to overlap due to
the (almost) degenerate masses for the $B$ mesons 
($m_{B^+} \simeq m_{B^0}$), and, for the $\bar B$ mesons 
($m_{B^-} \simeq m_{\bar {B^0}}$) in the medium, 
the negligible mass difference arising due to the very small 
difference in their vacuum masses. 
The threshold density above which the decay widths become nonzero
(with the center of mass momentum, $|{\bf p}|$ attaining a nonzero value) 
is observed to be higher for nuclear matter
(shown in (a)) as compared to hyperonic matter (shown in (c)).
For the case of isospin asymmetric nuclear
matter, the partial decay widths are seen to remain zero 
even upto a density of 6$\rho_0$ (shown in (b)), 
whereas for isospin asymmetric hyperonic matter (shown in (d)),
the threshold densities for the decay channels (i) and (ii)
are observed to be around 4.9$\rho_0$ and 3.9$\rho_0$ respectively. 
 }
\label{upsln1sdw}
\end{figure}

\begin{figure}
\vskip -0.9in
\includegraphics[width=18cm,height=17cm]{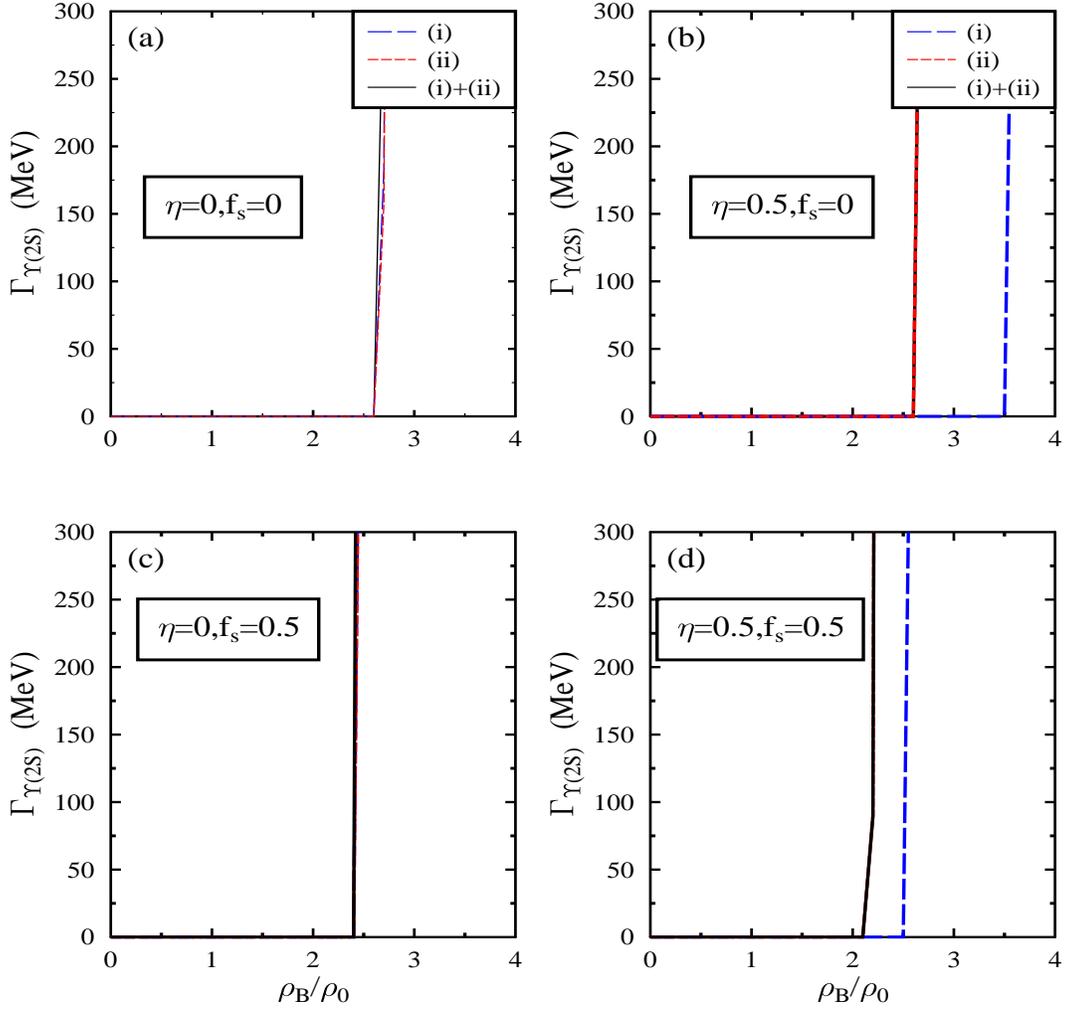}
\vskip -0.9in
\caption{(Color online) The partial decay widths of $\Upsilon (2S)$
calculated using the present model for composite hadrons,
to (i) $B^+B^-$, (ii) $B^0\bar {\rm B^0}$ and (iii) the sum of the 
two channels ((i)+(ii)) in the isospin symmetric (asymmetric)
nuclear matter ($f_s$=0) and strange hadronic matter ($f_s$=0.5). 
The channels (i) and (ii) overlap in the isospin symmetric ($\eta$=0) 
nuclear and hyperonic matter (shown in (a) and (c)) due to the mass 
of the $B^+$ ($B^-$) (almost) coinciding with the mass of the 
$B^0$ ($\bar {B^0}$).
In the symmetric (asymmetric) matter, the thresold densities above which 
the decay channels for (i) and (ii) are non zero, are observed
to be higher for nuclear matter, shown in (a) ((b)), 
as compared to hyperonic matter, shown in (c) ((d)).
 }
\label{upsln2sdw}
\end{figure}

\begin{figure}
\vskip -0.9in
\includegraphics[width=18cm,height=17cm]{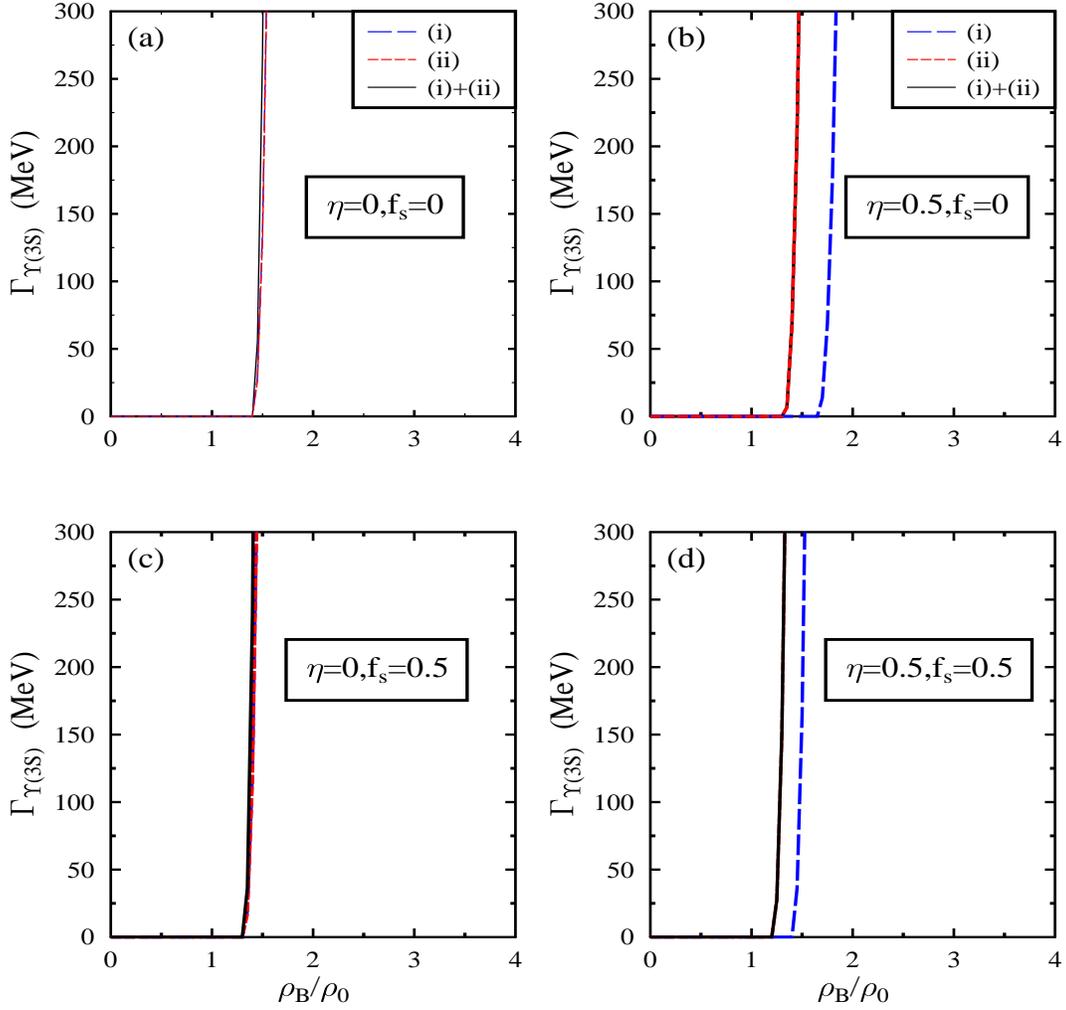}
\vskip -0.9in
\caption{(Color online) The partial decay widths of $\Upsilon (3S)$
calculated using the present model for composite hadrons,
to (i) $B^+B^-$, (ii) $B^0\bar {\rm B^0}$ and (iii) the sum of the 
two channels ((i)+(ii)) in the isospin symmetric (asymmetric)
nuclear matter and strange hadronic matter. The overlap of the 
values of the decay widths in channels (i) and (ii) in isospin
symmetric matter ($\eta$=0) in nuclear and hyperonic matter
(shown in (a) and (c) respectively),
are due to the mass of $B^+$ ($B^-$) being (almost) identical 
with the mass of $B^0$ ($\bar {B^0}$) in the medium, with the
negligible difference in the mass arising due to the very small
difference in the vacuum masses. In panels (b) and (d), the 
decay widths are plotted for the isospin asymmetric nuclear
and hyperonic matter. In both the symmetric and asymmetric
matter, the decay widths are observed to be nonzero above 
a threshold value, which is observed to be larger for 
the case of nuclear matter (shown in (a) and (b)) 
as compared to hyperonic matter (shown in (c) and (d)). 
}
\label{upsln3sdw}
\end{figure}

\begin{figure}
\vskip -1.1in
\includegraphics[width=16cm,height=16cm]{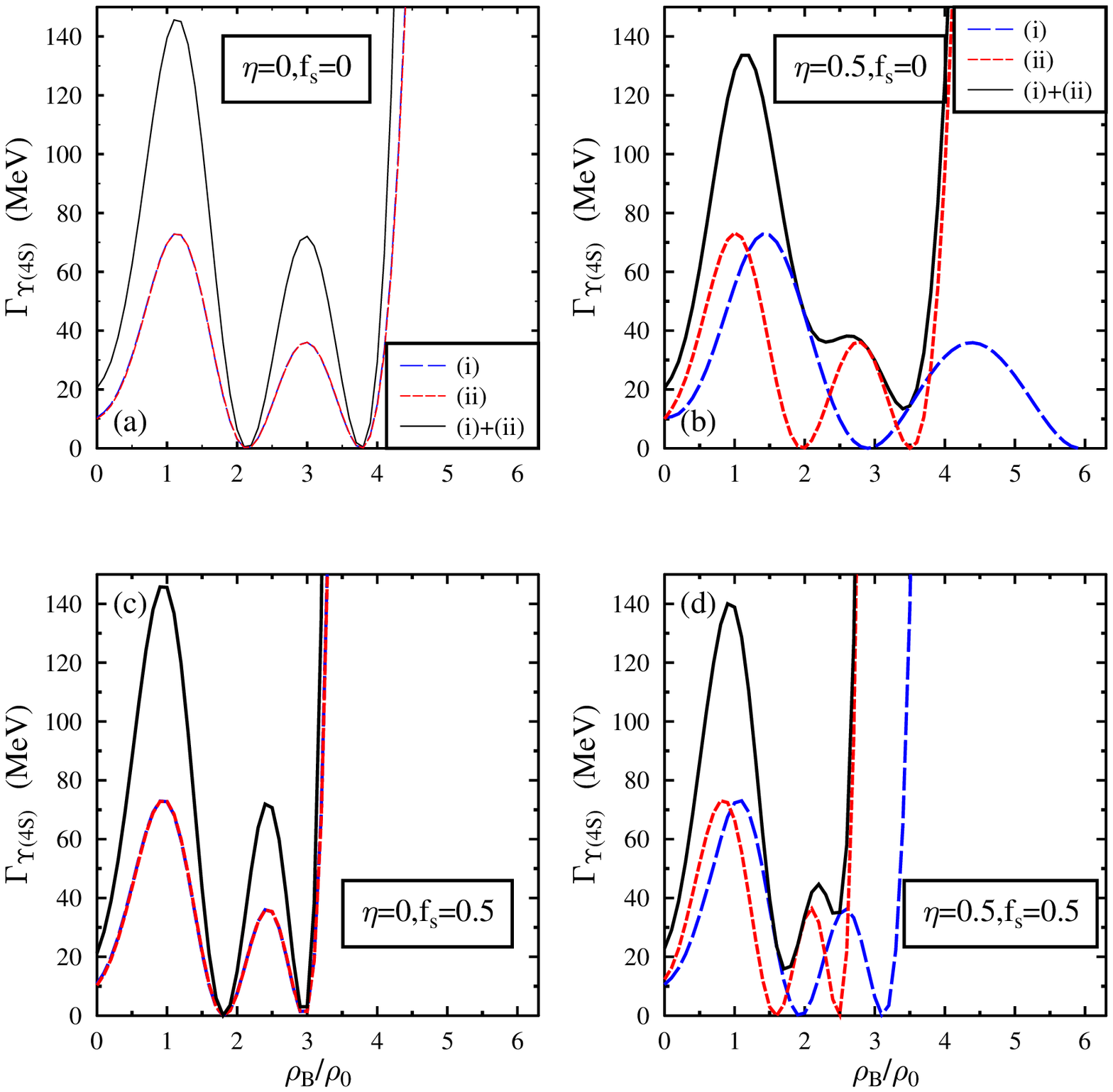}
\vskip -0.9in
\caption{(Color online) The partial decay widths of $\Upsilon (4S)$
calculated using the present model for composite hadrons,
to (i) $B^+B^-$, (ii) $B^0\bar {\rm B^0}$ and (iii) the sum of the 
two channels ((i)+(ii)) in the isospin symmetric (asymmetric)
nuclear matter and strange hadronic matter.
The decay widths for the channels 
(i) and (ii) are observed to overlap for the isospin symmetric ($\eta$=0)
for nuclear matter and hyperonic matter (as shown in (a) and
(c)) due to the mass of $B^+$ ($B^-$) to be close to the mass of
$B^0$ (${\bar {B^0}}$).
There is observed to be a rise with density followed by
a drop, with the decay widths reaching a value close to
zero (the node), at a value of around 2.1$\rho_0$ (1.8$\rho_0$)
for nuclear (hyperonic) matter. With further increase
in density, the value of the decay width is observed to
rise, reaching a peak, followed by a drop, leading again to 
a value close to zero (the second node) at a density of around 3.8$\rho_0$ 
(3$\rho_0$) for  nuclear (hyperonic) matter. With further increase
in density, there is observed to be a sharp rise in the
decay widths. In isospin asymmetric  ($\eta$=0.5) nuclear matter ($f_s$=0) 
shown in (b), the behaviour of the decay widths in both the channels 
remain similar, but the peaks and nodes
are observed to be at larger densities for channel (i) 
as compared to channel (ii). With the inclusion of hyperons in
the isospin asymmetric medium, these peaks and nodes 
are observed at smaller densities (as shown in (d)) as compared 
to asymmetric nuclear matter (as shown in (b)). 
}
\label{upsln4sdw}
\end{figure}

\section{In-medium masses of Open bottom mesons ($B$ and $\bar B$) 
and bottomonium states}

The partial decay widths of the bottomonium states, $\Upsilon (NS)$
($N=1,2,3,4$), to $B\bar B$ pair, in the hadronic medium, are calculated
in the present work. The medium modifications of these decay widths
arise due to the mass modifications of the decaying $\Upsilon$-state
and the open bottom ($B$ and $\bar B$) mesons.    
The in-medium masses of the heavy-light $B$ and $\bar B$ mesons 
are calculated within
an effective hadronic model, which is a generalization
of a chiral SU(3) model \cite{paper3}, 
based on the nonlinear realization of chiral symmetry 
\cite{weinberg,coleman1,coleman2,bardeen} 
and broken scale invariance \cite{hartree,kristof1,amdmeson}, 
to charm and bottom sectors,
so as to derive the interactions of these mesons with the hadronic medium.
The mass modifications of the open bottom mesons ($B$ and $\bar B$) arise from their interactions with the nucleons, 
hyperons and the scalar mesons in the strange hadronic matter
\cite{dpambmeson}. 
The medium modifications of the
bottomonium masses, on the other hand, arise from the
medium modifications of the gluon condensate in the hadronic medium.
The gluon condensate of QCD is simulated by a scale breaking term
\cite{paper3,sche1},
written in terms of a scalar dilaton field within the effective 
hadronic model \cite{amarvind,amcharmdw}. 
Matching the trace of the energy momentum tensor 
in QCD to that corresponding to the scale breaking term
in the effective hadronic model
gives the expression for the gluon condensate
in terms of the dilaton field
\cite{charmmass2,cohen,amarvind,amcharmdw}.
The medium 
modification of the gluon condensate in the hadronic matter 
is thus calculated from the modification of the
dilaton field, using which the in-medium masses
of the bottomonium states are calculated \cite{amdpbottomonium}.
Using a QCD sum rule approach \cite{kimlee,klingl}, 
the mass modifications of the
the charmonium states, were calculated
using the medium modifications
of the gluon condensates, obtained
from the medium change of the dilaton field 
within the effective hadronic model
\cite{charmmass2}.
The leading order perturbation calculations \cite{pes1}
and the framework of QCD sum rule 
yield the relation between medium modifications of the
heavy quarkonium states (charmonium and bottomonium states)
to the medium modifications of the gluon condensates \cite{leeko}.
The in-medium charmonium masses have been computed 
using the medium modifications of the gluon condensates 
calculated from the dilaton field in the hadronic medium,
within the effective hadronic model \cite{amcharmdw}. Using the 
medium changes of the masses of the charmonium states,
and, the $D$ and $\bar D$ mesons, the partial decay widths of the 
charmonium states to $D\bar D$ have been investigated 
using $^3P_0$ model \cite{amcharmdw} as well as using
the model for composite hadrons with quark and antiquark constituents
\cite{chmdwamspmwg}. 
In Ref. \cite{friman}, using a quark antiquark pair creation model,
namely $^3P_0$ model, the in-medium decay widths of the charmonium 
states, $\psi (3686)$, $\psi (3770)$, $\chi_{c0} (3417)$
and  $\chi_{c2} (3556)$ to $D\bar D$ pair were studied,
assuming the mass drops to be the same for $D$ as well as
$\bar D$ mesons in the medium and without accounting for
medium modifications of the masses of the charmonium sates.
There were observed to be nodes for the in-medium 
decay widths for $\psi(3686)$, $\psi(3770)$ as well as
$\chi_{c0} (3417)$, with decrease in the masses of 
$D$ and $\bar D$, whereas the decay width of the 
charmonium state $\chi_{c2} (3556)$ showed a monotonous 
increase with the drop in $D$ and $\bar D$ masses.
Using the mass modifications of the $D$ and $\bar D$ 
mesons as calculated in a chiral effective model,
also showed nodes for the in-medium decay widths
of the charmonium states, $\psi(3686)$ and $\psi(3770)$,
within the $^3P_0$ model, as well as, in the model
for composite hadrons, similar to as observed in
Ref. \cite{friman}. However, the decay width of 
$J/\psi \rightarrow D\bar D$ showed a monotonous increase
with the increase in density. The observed behaviour 
of the decay widths of the charmonium states to $D\bar D$ 
in the medium, is due to the competing effects
from a polynomial part and an exponential part
through the center of mass momentum, $|\bf p|$ of
the outgoing $D(\bar D)$ meson. Accounting for 
the effects of the medium modifications of the charmonium
states, was observed to decrease the contribution
from the exponential part, as the center of mass momentum,
$|\bf p|$ decreases, when the mass of the charmonium state
drops in the medium. With the mass modifications of the
charmonium states, as well as $D$ and $\bar D$ mesons,
there were no nodes observed upto a density of
6$\rho_0$. 
In the following section, using the model for composite hadrons,
we shall investigate the medium effects of the
bottomonium decay widths for the channel $\Upsilon \rightarrow 
B \bar B$, as arising from the medium modifications of the masses
of the $B$, $\bar B$ mesons \cite{dpambmeson}
as well as of the masses of the bottomonium states 
\cite{amdpbottomonium}.

\section{Results and Discussions}

In the present paper, we investigate the medium changes of the partial
decay widths of the bottomonium states ($\Upsilon(NS),N=1,2,3,4$)
to $B\bar B$ in hadronic matter, arising due to the modifications 
of the masses of these
hadrons calculated in an effective hadronic model.
The parameters chosen for the study of these decay widths
are as follows.
The masses of the $u$ and $d$ quarks are assumed 
to be 330 MeV and the mass of $b$ quark is taken 
to be 4180 MeV \cite{pdg2012} in the present work.
The values of the parameters $\lambda_1$ and  $\lambda_2$ 
are then calculated 
using equations (\ref{omega1}), (\ref{omega2}) and
(\ref{lambda12}), and, using the mass of the $B(\bar B)$ meson
(in vacuum) as 5279 MeV. These values turn out to be
0.1975 and 0.8025, respectively.

The parameters corresponding to the strengths of the harmonic oscillator 
wave functions for the $\Upsilon (NS)$ states are evaluated 
from the decay width $\Upsilon (NS) \rightarrow e^+e^-$,
given by the formula \cite{vanroyen,spmmesonspect} 
\begin{equation}
\Gamma_{\Upsilon (NS) \rightarrow e^+e^-}
=\frac{16 \pi \alpha^2}{9m_{\Upsilon (NS)}^2}
|\psi _{\Upsilon (NS)} (\bf 0)|^2,
\label{leptonicdw}
\end{equation}
where, $\alpha={1}/{137}$ is the fine structure constant,
$m_{\Upsilon (NS)}$ is the mass of $\Upsilon(NS)$ in vacuum, 
and $\psi _{\Upsilon (NS)}(\bf 0)$
is the wave function of the bottomonium state, $\Upsilon (NS)$
at the origin. Using the experimental values of the leptonic 
decay widths of 1.34 keV, 0.612 keV, 0.443 keV and 0.272 keV
for the $\Upsilon$-states, $\Upsilon (1S)$, $\Upsilon (2S)$,
$\Upsilon (3S)$ and $\Upsilon (4S)$ \cite{pdg2012},
we obtain the values of the harmonic oscillation strength, 
$R^{-1}_{\Upsilon(NS)}$ of the bottomonim states $\Upsilon (NS)$
to be 1309.2, 915.4, 779.75 and 638.6 MeV for $N=1,2,3,4$
respectively \cite{amdpbottomonium}.
For the study of the charmonium partial decay widths 
to $D\bar D$ pair in matter, the value of the 
wave function parameter, $R_D$ of the $D(\bar D)$ meson 
and the light quark pair production parameter, $\gamma$ 
were fitted from the decay widths of $\psi '' \rightarrow D\bar D$
and the partial decay widths of $\psi (4040)$ to $D\bar D$,
$D^* \bar D$, ${\bar D}^* D$ and $D^* {\bar D}^*$ 
\cite{leeko,chmdwamspmwg}.
The value of $R_D^{-1}$ was obtained as 310 MeV
\cite{leeko,chmdwamspmwg}.
In the present work, we assume the wave function parameter
for the $B(\bar B)$ meson, $R_B$ to be given in terms 
of the parameter $R_D$ of the $D(\bar D)$ meson wave function,
as $R_B= R_D (m_D/m_B)$, which yields the value of $R_B^{-1}$ to be 
875.6 MeV. The value of the strength of the light quark pair creation 
$\gamma_\Upsilon$ in the expression for the partial decay width
of the $\Upsilon$ state to $B\bar B$, is fitted from the vacuum
decay width of $\Upsilon (4S) \rightarrow B\bar B$.
The experimental values of the decay widths of
$\Upsilon(4S)\rightarrow B^+B^-$ and 
$\Upsilon(4S)\rightarrow B^0{\bar {B^0}}$
as 10.516 MeV and 9.984 MeV respectively, 
yield the value of $\gamma_\Upsilon$ to be 5.6.

We next calculate the decay widths of the 
$\Upsilon$ states to $B\bar B$ pair in matter and show 
the density dependence of these decay widths 
in figures \ref{upsln1sdw}, \ref{upsln2sdw}, \ref{upsln3sdw}, 
and \ref{upsln4sdw}, for $\Upsilon (1S)$,   $\Upsilon (2S)$,  
$\Upsilon (3S)$ and, $\Upsilon (4S)$, respectively.  
The effects of isospin asymmetry as well as strangeness
of the hadronic matter on these decay widths are also 
illustrated in these figures.
The partial decay widths for 
$\Upsilon (NS)\rightarrow B^0 \bar {B^0}$
and $\Upsilon (NS)\rightarrow B^+ {B^-}$ are given by   
equations (\ref{gammaupslnb0b0b}) and
(\ref{gammaupslnbpbm}), where $A^{\Upsilon (NS)}(|\bf p|)$
is given by equation (\ref{ap}). The decay widths thus have 
an exponential as well as polynomial dependence on the center of mass
momentum, $|\bf p|$. 
To understand the observed behaviour of
these decay wdiths in matter, as obtained in the present
investigation, it is useful to write these decay widths
in terms of the exponential and polynomial parts, as
$\Gamma (\Upsilon (NS)\rightarrow B\bar B )$
=$\exp (2C^F_{\Upsilon (NS)}|{\bf p}|^2)\times 
\Gamma_{polynomial} (\Upsilon (NS) \rightarrow B \bar B )$,
where $C^F_{\Upsilon (NS)}= (a_{\Upsilon (NS)} 
{b^2_{\Upsilon (NS)}} -R_B^2\lambda_2^2)$
is the coefficient of $|{\bf p}|^2$ in the exponential 
part of $A^{\Upsilon (NS)}(|{\bf p}|)$
given by equation (\ref{ap}). 
In figure \ref{upsln1sdw}, the partial decay widths 
of $\Upsilon (1S) \rightarrow B^+B^-$,  
$\Upsilon (1S) \rightarrow B^0 \bar {B^0}$, 
as well as the sum of these two channels are shown as
functions of the baryon density in units of the nuclear matter
saturation density, $\rho_0$. These are shown for the
isospin symmetric as well as asymmetric nuclear and hyperonic 
matter cases. The decay width for $\Upsilon (1S) \rightarrow B\bar B$
in symmetric nuclear matter is observed to be zero upto 
a density of about 5.2$\rho_0$, above which there is observed
to be a sharp rise with density. 
The sharp rise of this decay width with density can be
understood from the contributions of the exponential and
polynomial parts as follows.  
With the values of $R_{\Upsilon (1S)}$ and $R_B$ as already
mentioned to be $(1309.2\; {\rm MeV})^{-1}$ and 
$(875.6\; {\rm MeV})^{-1}$,
and the value of the parameter $\lambda_2$ as 0.8025, the values 
of $a_{\Upsilon(1S)}$ and $b_{\Upsilon(1S)}$ defined by
equation (\ref{abupsln}) are calculated, using which the value
of $C^F_{\Upsilon(1S)}$ is obtained as
$-0.15\times 10^{-6}$ MeV$^{-2}$. The value for the exponential
part of the decay width is observed to vary very less with
density (from a value of around 0.995
at a density of 5.2$\rho_0$ to about 0.48 at 6$\rho_0$).
In the polynomial part, the contributions are from the
first two terms of the expression for $A^{\Upsilon (NS)}(|{\bf p}|)$
given by equation (\ref{ap}), along with the $|{\bf p}|$ 
dependent terms (modulo the exponential part) multiplying 
$A^{\Upsilon (NS)}(|{\bf p}|)^2$ in the expressions 
for the decay widths of $\Upsilon (NS) \rightarrow B^0 {\bar {B^0}}$
and $\Upsilon (NS) \rightarrow B^+ B^-$ given by equations
(\ref{gammaupslnb0b0b}) and
(\ref{gammaupslnbpbm}) respectively.
For $\Upsilon (1S) \rightarrow B \bar B$, it is observed that
the second term within the square bracket in the expression
for $A^{\Upsilon (NS)}(|{\bf p}|)$ 
(with a value of 1.075) dominates over the 
first term ($|{\bf p}|$ dependent, which is negative and 
is of the order of $-0.124$ for a density of 6$\rho_0$).
Hence the value of the sum of the two terms within the
square bracket in the expression for 
$A^{\Upsilon (NS)}(|{\bf p}|)$ stays close to unity.
With increase in density, since the dependence on the exponential 
as well as the polynomial part of 
$A^{\Upsilon (NS)}(|{\bf p}|)$ are very small, 
the dependence of the decay widths of $\Upsilon (1S) \rightarrow
B\bar B$ is proportional to $|{\bf p}|^3 p_B^0 p_{\bar B}^0$,
which is observed as a sharp rise in the decay width 
of $\Upsilon (1S) \rightarrow B\bar B$. 
For symmetric nuclear matter, for $\rho_B=5.2 \rho_0$, 
the values of the polynomial part and the exponential part are
observed to be 24.3 MeV and 0.9946 (obtained from the value of
$|{\bf p}|$ as 129 MeV), with their product (the decay 
width of $\Upsilon (1S) \rightarrow B^+ B^-$) as 24.17 MeV. 
For a density of $5.24\rho_0$, the value of $|{\bf p}|$ as
343 MeV, gives the contributions from the polynomial and exponential
parts as 456 MeV and 0.965, giving the value of the decay width of 
$\Upsilon (1S) \rightarrow B^+B^-$ as 440 MeV. So there is observed
to be a sharp rise of the decay width with density in symmetric nuclear
matter. 
The threshold density above which
these decay widths become non-zero is observed to be smaller 
($\simeq$ 4.4$\rho_0$), with inclusion of hyperons in the medium,
as seen from panel (c) of figure \ref{upsln1sdw}. This is because
the masses of the $B(\bar B)$ mesons have a larger drop in nuclear matter
as compared to the mass drop in hyperonic matter. Also, the 
medium modifications of the open bottom mesons dominate over 
the modification of the $\Upsilon (1S)$, leading to a smaller value 
of the center of mass momentum, $|\bf p|$ in hyperonic matter,
as can be seen from the expression of $|\bf p|$ given by
equation (\ref{pb}). For symmetric hyperonic matter ($f_s$=0.5),
the contributions of the polynomial and the exponential
parts to the decay width are observed to be 60.74 MeV and
0.99 (with $|{\bf p}|$ as 177 MeV) leading to the 
partial decay width of $\Upsilon (1S) \rightarrow B^+ B^-$
as 60.13 MeV, and for a density of 4.55$\rho_0$, the value is
observed to increase to 630 MeV (with 661 MeV and 0.953 from the
polynomial and exponential parts of the decay width).
The isospin asymmetry in the hadronic matter 
leads to a smaller drop in the masses of the $B$ and $\bar B$
mesons, thus shifting the threshold density for the decay 
width to a larger value of density. For asymmetric nuclear matter,
the decay width remains zero even upto a density of around
6$\rho_0$, as can be seen from the panel (b) of
figure \ref{upsln1sdw}. With inclusion of hyperons in the medium,
the threshold densities for $\Upsilon (1S) \rightarrow B^+B^-$
and $\Upsilon (1S)\rightarrow B^0 \bar {B^0}$ are observed to
be around 4.9$\rho_0$ and 3.9$\rho_0$ respectively, for $f_s$=0.5,
as shown in panel (d) of figure \ref{upsln1sdw}.
A similar trend is observed for the in-medium partial
decay widths of the bottomonium states, $\Upsilon (2S)$
and $\Upsilon (3S)$, plotted in figures  \ref{upsln2sdw} 
and \ref{upsln3sdw} respectively. The thresold density 
above which the decay of $\Upsilon (2S)$ to $B\bar B$ 
becomes possible in the symmetric nuclear matter is around 
2.6$\rho_0$, which is observed to become smaller 
($\simeq$ 2.4$\rho_0$) when hyperons are included 
in the hadronic medium.
As has been observed for the case of the partial decay width
of $\Upsilon (1S) \rightarrow B\bar B$, for $\Upsilon (2S)$
decaying to $B\bar B$, there is oberved to be a sharp
rise in the decay width, 
again predominantly due to the polynomial part (proportional to 
$|{\bf p}|^3 p_B^0 p_{\bar B}^0$) multiplying the
$A^{\Upsilon (NS)}(|{\bf p}|)^2$ in the expression
for the decay widths of $\Upsilon (2S)\rightarrow B\bar B$.
This is again due to the reason that the contribution 
from $A^{\Upsilon (NS)}(|{\bf p}|)$ 
from the terms in the square bracket 
is observed to vary from 0.83 at 2.7$\rho_0$ 
(with a value of $|{\bf p}|$ to be 179 MeV)
to about 0.78 at 4$\rho_0$ (with a value of 
$|{\bf p}|$ to be 1121 MeV), and the exponential 
part is seen to vary from 0.983 at 2.7$\rho_0$ 
to 0.515 at 4$\rho_0$, due to the small value of the coefficient
$C^F_{\Upsilon (2S)}$ as $-0.264\times 10^{-6}$ MeV$^{-2}$.
For isospin asymmetric nuclear matter (with $\eta$=0.5), 
the threshold densities for the decay channels 
$\Upsilon (2S) \rightarrow B^+B^-$
and $\Upsilon (2S)\rightarrow B^0 \bar {B^0}$ are observed to
be 2.6$\rho_0$ and 3.5$\rho_0$ respectively, as shown 
in panel (b) of figure \ref{upsln2sdw}.
These values are observed to be modified to 2.1$\rho_0$ and 2.5$\rho_0$
for isospin asymmetric hyperonic matter (with $\eta$=0.5 and
$f_s$=0.5) as shown in panel (d) of figure \ref{upsln2sdw}.
From panels (a) and (c) of figure \ref{upsln3sdw}, 
the threshold densities 
for the decay process of $\Upsilon (3S) \rightarrow B \bar B$
are observed to be around 1.45$\rho_0$ and 1.35$\rho_0$ 
for symmetric nuclear matter and symmetric hyperonic matter.
The density dependence of the contribution from the exponential part 
of the decay width in symmetric nuclear matter  
is observed to be very small (0.966 at 1.45$\rho_0$
corresponding to $|{\bf p}|$ as 229 MeV,
to 0.79 at 1.75$\rho_0$ corresponding to $|{\bf p}|$ to be
around 652 MeV) and the value of the expression in the
square baracket of $A^{\Upsilon (NS)}(|{\bf p}|)$,
is observed to vary from 0.21 at 1.45$\rho_0$ to 
0.5 at 1.75$\rho_0$. Hence the density depedence
(through $|{\bf p}|$) of the decay widths of 
$\Upsilon (3S) \rightarrow B^0 {\bar {B^0}}$
and $\Upsilon (3S) \rightarrow B^+ B^-$ given by equations
(\ref{gammaupslnb0b0b}) and
(\ref{gammaupslnbpbm}) respectively, are due to the 
factor $|{\bf p}|^3 p_B p_{\bar B}$, multiplying the
$A^{\Upsilon (NS)}(|{\bf p}|)^2$ in the expression
for the decay widths of $\Upsilon (3S)\rightarrow B\bar B$.
This is the reason for the sharp rise of the
$\Upsilon (3S) \rightarrow B\bar B$ decay width
with density.
With inclusion of isospin asymmetry in the medium,
the densities above which the decay processes 
$\Upsilon (3S) \rightarrow B^+B^-$
and $\Upsilon (3S)\rightarrow B^0 \bar {B^0}$ become possible,
are 1.7$\rho_0$ and 1.35$\rho_0$ for nuclear matter,
as can be seen from the panel (b),
and 1.45$\rho_0$ and 1.25$\rho_0$ for hyperonic matter,
as can be seen from the panel (d) of
figure \ref{upsln3sdw}.

In figure \ref{upsln4sdw}, we show the density dependence of
the decay width $\Upsilon (4S) \rightarrow B\bar B$
for isospin asymmetric as well as isospin asymmetric 
nuclear and hyperonic matter. As has already been mentioned,
the production strength of the light quark-antiquark pair
for this decay, $\gamma_\Upsilon$, is fitted from the experimental
(vacuum) decay widths of $\Upsilon (4S) \rightarrow  
B^+B^-$ and $\Upsilon(4S)\rightarrow B^0{\bar {B^0}}$
of 10.516 MeV and 9.984 MeV respectively (the small difference
is due to the small difference in the vacuum masses of $B^+(B^-)$ 
and $B^0 (\bar {B^0})$). In isospin symmetric nuclear matter,
as can be seen from panel (a) of figure \ref{upsln4sdw},
the decay width of $\Upsilon (4S) \rightarrow B^+ B^-$
(which is almost identical to the decay width of
$\Upsilon (4S) \rightarrow B^0 {\bar {B^0}}$ in the
symmetric matter)
is observed to increase with density reaching a value of around
70 MeV at a density of around 1.1$\rho_0$. 
Above this value of density,
there is seen to be a drop in the decay width with further increase 
in density, and it is observed that
the decay width vanishes at a density of around 2.1$\rho_0$.
As the density is further increased, the decay width  is also seen to
increase, again reaching a maximum value of around 35 MeV at a density of
around 3$\rho_0$ for both the channels $\Upsilon (4S) \rightarrow B^+B^-$ 
and $\Upsilon (4S) \rightarrow B^0 \bar {B^0}$. 
There is again seen to be a drop and ultimately vanishing of
the decay width at a density of around 3.8$\rho_0$.
Above this density, there is observed to be a sharp rise
in the value of the decay widths. 
For symmetric nuclear matter, the contribution of the exponential
part of the decay width is observed to vary from 0.92 at zero density
to around 0.11 at density of 4$\rho_0$. The density dependence of the
part of decay width corresponding to the terms in the square bracket in 
the expression for $A^{\Upsilon (NS)}(|\bf p|)$ given by equation (\ref{ap}), 
is observed to be as follows. Its value of 0.045 at zero density,
is observed to increase slowly upto a value of 0.0466 at a density 
of around 0.5$\rho_0$, and decreases with further increase in 
density, becoming very small ($\sim 6.4\times 10^{-6}$ 
at around 2.1$\rho_0$). There is observed
to be a rise in its value with density upto a density of around 3$\rho_0$,
followed by a drop and again reaching a value of around 6.63
$\times 10^{-6}$ at around 3.8$\rho_0$. As the density is further
increased, there is observed to be a steady increase in its value
reaching a value of around 1.12 at a density of 5.9$\rho_0$.
The value of $|{\bf p}|$ is observed to increase from a value
of 323 MeV at zero density to around 1630 MeV at 4$\rho_0$.
The density dependence of the decay width is thus 
due to combined effects of the contributions from the
exponential term (decreasing with density),
the factor $|{\bf p}|^3 p_B p_{\bar B}$, multiplying the
$A^{\Upsilon (NS)}(|{\bf p}|)^2$ in the expression
for the decay width of $\Upsilon (4S)\rightarrow B\bar B$,
as well as the contribution from the terms within 
the square bracket in the expression of 
$A^{\Upsilon (NS)}(|{\bf p}|)$ given by equation (\ref{ap}).
The observed behaviour of the decay width for the process
$\Upsilon (4S)\rightarrow B\bar B$
in symmetric nuclear matter is thus dominantly due to
the contribution of the polynomial part of the decay width,
arising from the terms within the square bracket 
in the expression of $A^{\Upsilon (NS)}(|{\bf p}|)$.
The behaviour of the decay width of $\Upsilon (4S) \rightarrow B\bar B$
with density, is observed to remain similar, with inclusion of hyperons 
in the medium, as shown in panel (c) of the figure \ref{upsln4sdw}, 
but the densities at which the decay widths
reach maximum values (at densities of 0.9$\rho_0$ and 2.5$\rho_0$) 
and vanish (at densities of 1.8$\rho_0$ and 3$\rho_0$) are observed 
to be smaller than those for isospin symmetric nuclear matter. 
The isospin asymmetry effects on the partial decay widths of 
the process $\Upsilon (4S) \rightarrow B\bar B$ in nuclear
matter and hyperonic matter are also
shown in panels (b) and (d) of figure \ref{upsln4sdw}.
In asymmetric nuclear matter, the difference in the
decay widths of $\Upsilon (4S) \rightarrow B^+B^-$ 
and $\Upsilon (4S) \rightarrow B^0 \bar {B^0}$
are quite pronounced at high densities as can be seen 
in panel (b) of figure \ref{upsln4sdw}, even though the
qualitative features remain the same as 
for symmetric nuclear matter. The decay width
$\Upsilon (4S) \rightarrow B^+B^-$
($\Upsilon (4S) \rightarrow B^0 \bar {B^0}$)
is observed to rise with increase in density upto a density of
around 1.5$\rho_0$ ($\rho_0$) followed by a drop with further
increase in density upto around 2.9$\rho_0$ (2$\rho_0$) 
when the decay width is observed to vanish. 
Above the density at which there is a node, i.e., the decay width 
attains zero value,
with further increase in density, the decay width is observed 
to again attain a maximum at around 4.4$\rho_0$ (2.8$\rho_0$) 
followed by another node at density of 6$\rho_0$ (3.5$\rho_0$) 
for the decay process $\Upsilon (4S) \rightarrow B^+B^-$
($\Upsilon (4S) \rightarrow B^0 \bar {B^0}$).
The effects of isospin asymmetry are observed to be
smaller for hyperonic matter (shown in (d) of figure \ref{upsln4sdw}), 
as compared to the case of nuclear matter (shown in (b)), 
even though qualitative bahaviour
remains the same. The values of densities at which the decay
widths of $\Upsilon (4S) \rightarrow B^+B^-$
($\Upsilon (4S) \rightarrow B^0 \bar {B^0}$) have maxima
are 1.2$\rho_0$ (0.8$\rho_0$) and 2.6$\rho_0$ 
(2.1$\rho_0$) and the values where nodes 
are observed are 1.9$\rho_0$ (1.6$\rho_0$) 
and 3.2$\rho_0$ (2.5$\rho_0$) respectively.

As has already been mentioned the in-medium decay widths
of the bottomonium states to $B\bar B$, 
have contributions of a polynomial part and an exponential part,
in terms of the center of mass momentum, $|\bf p|$ defined through
equation ({\ref{pb}}). This can be
seen from the expressions of the 
partial decay widths of $\Upsilon (NS) \rightarrow B^0 {\bar {B^0}}$
and  $\Upsilon (NS) \rightarrow B^+ B^-$ given by
equations (\ref{gammaupslnb0b0b}) and
(\ref{gammaupslnbpbm}), with $A^{\Upsilon (NS)}(|\bf p|)$
as in equation (\ref{ap}). 
For the decay of $\Upsilon (1S)$, $\Upsilon (2S)$ and 
$\Upsilon (3S)$, there is observed to be a sharp rise
in the decay width with density, predominantly due to the
factor $|{\bf p}|^3 p^0_B p^0_{\bar B}$
multiplied to $A^{\Upsilon (NS)}(|{\bf p}|)^2$ in the expressions
of the partial decay widths of $\Upsilon (NS) 
\rightarrow B^0 {\bar {B^0}}$
and  $\Upsilon (NS) \rightarrow B^+ B^-$. For the decay of
$\Upsilon (NS) \rightarrow B\bar B$, $N=1,2,3$, 
the effects of density arising from the exponential part as well as 
from the expression within the square bracket in
the expression for $A^{\Upsilon (NS)}(|\bf p|)$ are 
observed to be very small.
For the partial decay width of $\Upsilon(4S)
\rightarrow B\bar B$ there is seen to be an initial rise
in the decay width with increase in density, followed by a decrease
with further  increase in the density leading to a node.
As the density is further increased there is a drop 
again after initial increase in the decay width leading to
observation of a second node. 
The observed behaviour of the decay width of
$\Upsilon (4S) \rightarrow B\bar B$ is dominantly due to the
behaviour of the contribution of the terms within the 
square bracket of the expression for 
$A^{\Upsilon (NS)}(|\bf p|)$.
In the present investigation of 
the in-medium decay widths of the bottomonium
states, we have accounted for the medium modifications
of the masses of the bottomonium states as well as 
the open bottom mesons ($B$ and $\bar B$ mesons). 
The finding of nodes for $\Upsilon (4S) \rightarrow B\bar B$
mesons should have observable consequences on the production
of the hidden as well as open bottom mesons in heavy ion
collision experiments.
  
\section{Summary}
In the present paper, we have investigated the 
partial decay widths of $\Upsilon \rightarrow B\bar B$
within a field theoretical model of composite hadrons.
The medium modifications of these decay widths arise
from the medium changes of the masses of
the bottomonium and open bottom mesons. The masses 
of the $B$ and $\bar B$ mesons, calculated in an 
effective chiral model, arise due to their 
interactions with the baryons and the scalar mesons.
On the other hand, the mass modifications of the bottomonium
states in the hadronic medium are due to the medium modification
of a scalar dilaton field, which is introduced in the
effective hadronic model to simulate the gluon condensates
of QCD. The effects of isospin asymmetry, strangeness 
on the bottomonium decay widths for decays
$\Upsilon (NS) \rightarrow B \bar B$, $N=1,2,3,4$, 
have also been studied in the present work. 
The density effects are seen to be the dominant medium
effects which should have observable consequences
from the dense hadronic matter created 
in heavy ion collision experiments.  
The decay width of $\Upsilon (4S) \rightarrow B\bar B$
is observed to show nodes at specific densities,
similar to what was observed for the partial decay widths
of the excited charmonium states to $D\bar D$ pair
within $^3P_0$ model, as well as, within a field theoretical
model of composite hadrons with quark and antiquark constitutents.
The isospin asymmetry effects are observed to be quite pronounced
at high densities, leading to quite different values for 
the decay widths of $\Upsilon (4S) \rightarrow B^+B^-$
and $\Upsilon (4S) \rightarrow B^0 {\bar {B^0}}$.
These isospin asymmetry effects at high densities 
on these partial decay widths should show in the
asymmetric heavy ion collision experiments in CBM planned at FAIR.
However, the study of the bottomonium states will require 
access to energies higher than the energy regime planned at CBM 
experiment. The density effects on the partial decay widths 
$\Upsilon \rightarrow B\bar B$ 
should show up in observables, e.g., the yield of the bottomonium
states and the open bottom mesons, as well as,
in the dilepton spectra at the Super Proton Synchrotron 
(SPS) energies.

{}
\end{document}